\documentclass[12pt, draftclsnofoot, onecolumn]{IEEEtran}

\usepackage{amsmath}

\usepackage{amsmath, mathrsfs}
\usepackage{amsfonts}
\usepackage{amssymb}
\usepackage{graphicx}
\usepackage{color}
\usepackage{multirow}
\usepackage{graphicx}
\usepackage{stmaryrd}
\usepackage{cool}
\usepackage[keeplastbox]{flushend}
\allowdisplaybreaks

\usepackage{amsfonts}
\usepackage{amssymb}
\usepackage{graphicx}
\usepackage{color}
\usepackage{multirow}
\usepackage{graphicx}

\usepackage{caption}
\usepackage{subcaption}
\usepackage[numbers,sort&compress]{natbib}

\newcommand{\subparagraph}{}
\usepackage[compact]{titlesec}
\titlespacing*{\section}{15pt}{1\baselineskip}{0.9\baselineskip}

\allowdisplaybreaks

\newcommand{\myhash}{%
	{\settoheight{\dimen0}{C}\kern-.05em\, \resizebox{!}{\dimen0}{\raisebox{\depth}{\#}}}}

\usepackage{multicol}


\usepackage{mathbbol}

\def\mindex#1{\index{#1}}

\def\sq{\hbox{\rlap{$\sqcap$}$\sqcup$}}
\def\qed{\ifmmode\sq\else{\unskip\nobreak\hfil
\penalty50\hskip1em\null\nobreak\hfil\sq
\parfillskip=0pt\finalhyphendemerits=0\endgraf}\fi\medskip}

\long\def\defbox#1{\framebox[.9\hsize][c]{\parbox{.85\hsize}{%
\parindent=0pt
\baselineskip=12pt plus .1pt      
\parskip=6pt plus 1.5pt minus 1pt 
 #1}}}

\long\def\beginbox#1\endbox{\subsection*{}%
\hbox{\hspace{.05\hsize}\defbox{\medskip#1\bigskip}}%
\subsection*{}}

\def\endbox{}

\newsavebox{\junk}
\savebox{\junk}[1.6mm]{\hbox{$|\!|\!|$}}

\def\sfH{{\sf H}}

\def\bfmath#1{{\mathchoice{\mbox{\boldmath$#1$}}%
{\mbox{\boldmath$#1$}}%
{\mbox{\boldmath$\scriptstyle#1$}}%
{\mbox{\boldmath$\scriptscriptstyle#1$}}}}

\def\bfmY{\bfmath{Y}}

\def\bfmhhaY{\bfmath{\hhaY}} 
\def\bfmhhaY{\hbox to 0pt{$\widehat{\bfmY}$\hss}\widehat{\phantom{\raise 1.25pt\hbox{$\bfmY$}}}}

\def\til={{\widetilde =}}

 \def\FRAC#1#2#3{\genfrac{}{}{}{#1}{#2}{#3}}

\def\ddtp{{\mathchoice{\FRAC{1}{d^{\hbox to 2pt{\rm\tiny +\hss}}}{dt}}%
{\FRAC{1}{d^{\hbox to 2pt{\rm\tiny +\hss}}}{dt}}%
{\FRAC{3}{d^{\hbox to 2pt{\rm\tiny +\hss}}}{dt}}%
{\FRAC{3}{d^{\hbox to 2pt{\rm\tiny +\hss}}}{dt}}}}

\def\average#1,#2,{{1\over #2} \sum_{#1}^{#2}}

\def\eye(#1){{\bf(#1)}\quad}

\newtheorem{theorem}{{\bf Theorem}}

\newtheorem{remark}{{\bf Remark}}

\def\eq#1/{(\ref{e:#1})}

\newcommand{\beqn}[1]{\notes{#1}%
\begin{eqnarray} \elabel{#1}}

\newcommand{\eeqn}{\end{eqnarray} }

\newcommand{\beq}[1]{\notes{#1}%
\begin{equation}\elabel{#1}}

\newcommand{\eeq}{\end{equation}}

\def\bdes{\begin{description}}
\def\edes{\end{description}}

\newcounter{rmnum}

\newcounter{anum}

{\end{list}}

\def\ass(#1:#2){(#1\ref{#1:#2})}

\def\ritem#1{
\item[{\sf \ass(\current_model:#1)}]
}

\newenvironment{recall-ass}[1]{%
\begin{description}
\def\current_model{#1}}{
\end{description}
}

\newfont{\bb}{msbm10 scaled 1100}
\newcommand{\CC}{\mbox{\bb C}}
\newcommand{\PP}{\mbox{\bb P}}
\newcommand{\RR}{\mbox{\bb R}}

\newcommand{\EE}{\mbox{\bb E}}

\newcommand{\dv}{{\bf d}}

\newcommand{\hv}{{\bf h}}

\newcommand{\qv}{{\bf q}}

\newcommand{\vv}{{\bf v}}
\newcommand{\xv}{{\bf x}}
\newcommand{\yv}{{\bf y}}
\newcommand{\zv}{{\bf z}}

\newcommand{\Hm}{{\bf H}}

\newcommand{\Nm}{{\bf N}}

\newcommand{\Pm}{{\bf P}}

\newcommand{\Tm}{{\bf T}}

\newcommand{\Ym}{{\bf Y}}

\newcommand{\Ac}{{\cal A}}
\newcommand{\Bc}{{\cal B}}
\newcommand{\Cc}{{\cal C}}

\newcommand{\Lc}{{\cal L}}

\newcommand{\Nc}{{\cal N}}

\newcommand{\Sc}{{\cal S}}

\newcommand{\Xc}{{\cal X}}

\newcommand{\Pim}{\hbox{\boldmath$\Pi$}}

\newcommand{\herm}{{\sf H}}

\newcommand{\SIR}{{\sf SIR}}

\usepackage{hyperref}
\hypersetup{
    bookmarks=true,         
    unicode=false,          
    pdftoolbar=true,        
    pdfmenubar=true,        
    pdffitwindow=false,     
    pdfstartview={FitH},    
    pdfnewwindow=true,      
    colorlinks=true,       
    linkcolor=red,          
    citecolor=green,        
    filecolor=blue,      
    urlcolor=blue           
}

\usepackage{tikz}
\usepackage{pgfplots,tikz-3dplot}
\pgfplotsset{compat=newest}
\usetikzlibrary{patterns}
\usetikzlibrary{positioning}
\usetikzlibrary{datavisualization}
\usetikzlibrary{datavisualization.formats.functions}
\usetikzlibrary{backgrounds}
\usetikzlibrary{shapes,snakes}

\usepgfplotslibrary{external} 
\usepackage{float}
\usepackage{afterpage}
\usepackage{balance}

\usepackage{stackengine}

\def\herm{{\sfH}}

\newcommand{\Dd}{{\mathrm{d}}}

\makeatletter
\newcommand{\ssymbol}[1]{^{\@fnsymbol{#1}}}
\makeatother

\begin{document}

\title{Achieving Spatial Scalability for Coded Caching over Wireless Networks}

\author{Mozhgan~Bayat,
	Ratheesh~K.~Mungara,
	and~Giuseppe~Caire
	\thanks{The authors are with the Communications and Information Theory Group, Technische Universit\"{a}t Berlin, 10623 Berlin, Germany. E-mails: \{bayat, mungara, caire\}@tu-berlin.de.}
}

\maketitle

\begin{abstract}
The coded caching scheme proposed by Maddah-Ali and Niesen 
considers the delivery of files in a given content library to users through a 
deterministic error-free network where a common multicast message is sent to all users at a fixed rate, independent 
of the number of users.  
In order to apply this paradigm to a wireless network, it is important to make sure that the common multicast rate 
does not vanish as the number of users increases.  
This paper focuses on a variant of coded caching successively proposed for the so-called {\em combination network}, 
where the multicast message is further encoded by a {\em Maximum Distance Separable} (MDS) 
code and the MDS-coded blocks are simultaneously transmitted from different {\em Edge Nodes} (ENs) (e.g., 
base stations or access points).  Each user is equipped with multiple antennas and can 
select to decode a desired number of EN transmissions, while either nulling of treating as noise 
the others, depending on their strength. The system is reminiscent of the so-called {\em evolved Multimedia Broadcast Multicast Service} (eMBMS), 
in the sense that the fundamental underlying transmission mechanism is multipoint multicasting, where each user can independently and individually 
(in a user-centric manner) decide which EN to decode, without any explicit association of users to ENs.  
We study the performance of the proposed system when users and ENs are distributed according to 
homogeneous Poisson Point Processes in the plane and the propagation is affected by Rayleigh fading and distance dependent pathloss. 
Our analysis allows the system optimization with respect to the MDS coding rate. Also, we show that the proposed 
system is fully scalable, in the sense that it can support an arbitrarily large number of users, while maintaining 
a non-vanishing per-user delivery rate. 
\end{abstract}


\section{Introduction}

With the forthcoming explosion of data traffic generated by the wireless Internet, it is important to consider schemes that 
take advantage of the user data consumption patterns. In particular, multimedia on-demand content delivery (especially video)
is by far the most bandwidth consuming and rapidly growing application. Users typically make requests from some pre-existing video database/server (e.g., Netflix, YouTube, Hulu, Amazon Prime). This means that the user requests are highly redundant. 
In LTE, the {\em evolved Multimedia Broadcast Multicast Service} (eMBMS) standard 
was designed to simultaneously multicast popular common contents (e.g., live TV channels) 
to users in a wide area, by exploiting simultaneous transmission from multiple base stations. 
On the other hand, in the case of on-demand content delivery,  the strongly asynchronous 
request patterns prevent from simply taking advantage of the broadcast nature of the wireless medium
and transmit the same few files to a very large number of users. 
In order to take advantage of such {\em asynchronous content reuse}, wireless edge caching has been widely advocated in several different contexts 
(see for example the seminal work in \cite{GSDMC:12,golrezaei2013femtocaching} and the stream of more recent works as in \cite{liu2016caching,paschos2016wireless,sengupta2016cloud,poularakis2016complexity}).

In this work, we focus in particular on the elegant and information-theoretic near-optimal scheme known as {\em coded caching}, introduced by Maddah-Ali and Niesen in \cite{maddah2014fundamental,maddah2015decentralized}. The scheme was originally 
proposed for an idealized {\em single bottleneck network}, where a server is connected to $K$ users through a shared error-free link of given capacity. The scheme of \cite{maddah2014fundamental,maddah2015decentralized} is based on two distinct phases: cache placement (or pre-fetching) and coded delivery. 
In the pre-fetching phase, to be performed before the network run time (e.g., in off-peak hours), the users cache segments of the library files 
either in a specified manner (centralized scheme \cite{maddah2014fundamental}) or  at random (decentralized scheme \cite{maddah2015decentralized}). 
Then, during the network runtime, users make requests  and the server responds with the transmission of a multicast coded message, 
computed such that each user can retrieve its requested file from the common multicast message and its own cache content. 
For the single bottleneck link network, the  coded caching scheme (with an improved delivery that takes into account possibly 
repeated multiple requests) has been shown to be strictly optimal over all schemes with uncoded pre-fetching, i.e., when the users can cache segments of 
the library files and not general functions thereof \cite{yu2017exact}.
In the recent work \cite{caching-delivery-separation-ISIT}, general network topologies are considered and it is shown that 
the caching and delivery scheme of  \cite{maddah2014fundamental,yu2017exact}, designed independently of the underlying network, 
yields optimal performance within a bounded gap on any network, thus assessing a sort of ``separation theorem'' between caching and delivery. 

Based on the coded caching idea, several other network topologies have been studied. A particularly relevant topology for 
the present paper is the so-called {\em combination network}, formed by the server, a layer of $N_E$ intermediate nodes acting as relays, and the users.
There are exactly $K = {N_E \choose L}$ users, each of which is connected to a distinct subset of  $L$ out of $N_E$ relays (from which the name ``combination'' network) \cite{ji2015caching,zewail2017coded,wan2017caching,wan2017novel,mital2017coded}. 
Both the single bottleneck link network and the combination network are deterministic error-free models. 
On the other hand, in order to apply such caching paradigms to an actual wireless network, it is important to ensure  that the content 
delivery rate does not vanish as the number of users become large. This represents a challenge in
the presence of small-scale fading and distance dependent pathloss. 
We call {\em scalable} a scheme achieving a per-user delivery rate that tends to a positive limit as $K \rightarrow \infty$. 
In particular, for networks with constant user spatial density (users per unit area), $K \rightarrow \infty$ implies that the network coverage area 
must grow linearly with $K$. In this case, a scalable scheme shall be referred to as {\em spatially scalable}, 
since the scheme can provide a fixed positive delivery rate for an arbitrarily large coverage area. 

The goal of this work is to propose and analyze a spatially scalable scheme for on-demand content delivery based on coded caching. 
We consider a scenario motivated by eMBMS, where the content server is connected through high-capacity backhaul links to $N_E$ wireless transmitters
(e.g.,  base stations, access points, or remote radio heads), here referred to as {\em Edge Nodes} (EN) \cite{sengupta2016cloud}.  
We argue that the combination network forms a good paradigm 
for a more general wireless network where any user can choose to receive data from $L$ ENs in a completely autonomous user-centric way. 
In fact, since the combination network must serve all possible ``user types'' for all possible ${N_E \choose L}$ connections, 
it follows that the users can roam free over the coverage area, and receive data as long as they can decode the messages of any subset of  
$L$ ENs  \cite{ji2015caching}.  
 
In coded caching, the common coded multicast message must be delivered to all users at a common rate, such that the 
system bottleneck is given by the worst-case user. 
It is well-known that such common broadcast rate vanishes in the presence of independent Rayleigh fading across the users  $K \rightarrow \infty$.
In \cite{ngo2017scalable} it is shown that transmitting with a multiantenna base station with growing 
number of antennas at least as $\log K$ yields a scalable system.
Nevertheless, such a single base station multi-antenna approach  will not be effective against the pathloss, 
since all antennas are concentrated at the base station site.  Therefore, the approach of \cite{ngo2017scalable}  is not
spatially scalable. In contrast, in the proposed system, we consider spatially distributed single-antenna ENs, and users equipped with multiple antennas.\footnote{Typical hand-held devices operating at conventional cellular and WiFi frequency bands
cannot host a large number of antenna elements. Nevertheless, for a relevant scenario we may think of a vehicular network
where users are cars, which can host relatively large arrays, or a mmWave network, where a large number of antenna elements can be accommodated in a small form factor even on hand-held devices.} 
Using stochastic geometry, we show that the proposed system yields spatially scalable per user throughput in the presence of both small-scale Rayleigh 
fading and distance dependent pathloss.  In addition,  the proposed system is extremely simple, since no coordination among users and between 
the users and the ENs is required,  and no joint signal space precoding across the ENs is used.

In current wireless technology, techniques that make use of spatially distributed antennas with various degrees of 
joint processing at the signal level are generally referred to as \textit{Cloud radio access network} (C-RAN). In these schemes, 
ENs are connected to a centralized cloud processor. In full joint-processing C-RAN architectures, 
the cloud processor implements the baseband processing functionalities, including multiuser MIMO precoding \cite{simeone2016cloud, C-RAN-compression, park2014fronthaul}. In the context of caching, \textit{Fog Radio Access Networks} (F-RANs) take a mixed approach between 
the C-RAN paradigm and {\em edge processing}, by equipping the ENs with caching capabilities \cite{peng2016fog, park2013joint, tandon2016cloud}.
In contrast to such cache-aided F-RAN systems, we consider that caching is performed in the user devices and, as already said, we assume
that the backhaul links do not represent the system bottleneck, which is typically the case in eMBMS applications. 
Furthermore, in our scheme no joint baseband processing of the ENs is required. 
In fact, each EN simply broadcasts a network-coded message at fixed rate, without need of {\em transmitter channel state information}.

\section{System model}
\label{sec:SystemModel}

We consider a wireless network with $N_E$ single-antenna ENs and $K$ users, each equipped with an $n_r$-antenna array. 
A server is connected to the ENs via an error-free backhaul data network, of very high capacity.
The system bottleneck is, realistically,  the wireless segment between the ENs and the users. 
The server has access to a content library with $N$ files represented by $\mathcal{F} = \{W_1,W_2,..., W_N\}$, each of which has a size of $F$ bits. 
Each user has a cache memory of capacity $MF$ bits. Since the backhaul network has very high capacity, the ENs do not need  cache memory since whatever they have to transmit can be provided to them by the server. Following the paradigm of coded caching, we consider the pre-fetching phase and the delivery phase. 
The pre-fetching phase is identical to the optimal centralized strategy for the single bottleneck link network. This is recalled here for the sake of completeness. 
For simplicity, assume that $t = MK/N$ is an integer (if not, a memory-sharing argument is used to multiplex between points for which this 
parameter is an integer \cite{maddah2014fundamental}). 
Each file $W_i$ is partitioned into segments\footnote{For an integer $n$ we let $[n] = \{1,2, \ldots, n\}$.} 
$\{W_{i}^T : T \subseteq [K], |T| = t \}$ and each user $k \in [K]$ caches all and only the segments 
$W_{i}^T$ for all the subsets $T$ such that $T \ni k$. It is easy to check that  the cache of each user must have size
	\begin{equation}
	N \frac{F}{{K \choose t}} {K - 1 \choose t - 1} = MF.
	\end{equation}

In the delivery phase, given a demand vector $\dv \in [N]^K$, the server computes the codeword $X$ obtained from the concatenation of 
blocks $X_S$ for all $S \subseteq [K], |S| = t+1$. Each such block is given by 
	\begin{equation}  \label{ziominchia}
X_S = \bigoplus_{k \in S} W_{d_k}^{S\setminus\{k\}},
\end{equation}
where we notice that for any $S$ of size $t + 1$ and $k \in S$, the set $S \setminus\{k\}$ is a subset of size $t$, such that 
$W_{d_k}^{S\setminus\{k\}}$ are effectively segments of the file desired by user $k$ and cached at all users $j \in S$, $j \neq k$. 
The multicast codeword defined in \eqref{ziominchia} corresponds to the original scheme proposed in \cite{maddah2014fundamental}. A slight modification of
this scheme yields given in \cite{yu2017exact} a better delivery rate when $K < N$ and therefore some users necessarily request the same file. 
We hasten to say that all the results in this paper extend immediately to this improved scheme. The two delivery schemes for the worst-case 
demand coincide when $K \geq N$, since the worst case demand (i.e., the demand configuration that requires the longest multicast codeword) contains
all distinct demands. The overall transmission length is given by 
	\begin{equation}
\Lc(X) = \frac{F}{{K \choose t}} {K  \choose t + 1}  =  \frac{F K (1- \mu)}{1 + K \mu},
	\end{equation}
where $\mu = M/N$ is the fractional cache memory. 

In our scheme, $X$ is divided into $L$ blocks of equal size, for some integer $L \leq N_E$ 
and {\em Maximum Distance Separable} (MDS) coding is applied to append $N_E - L$ parity blocks. 
Each MDS-coded block is treated as a symbol over a large finite field, with length in bits given by  $\frac{1}{L} \Lc(X)$. 
The resulting $N_E$ MDS-coded blocks are sent to the $N_E$ ENs, such that each EN is associated to a distinct MDS-coded block. 
Eventually,  the blocks are transmitted through an appropriate physical-layer (PHY) coded modulation scheme and 
sent in parallel, on the same time-frequency slot, from the different ENs. 

\begin{figure}[t]
	\centering
	\includegraphics[width=0.45\textwidth]{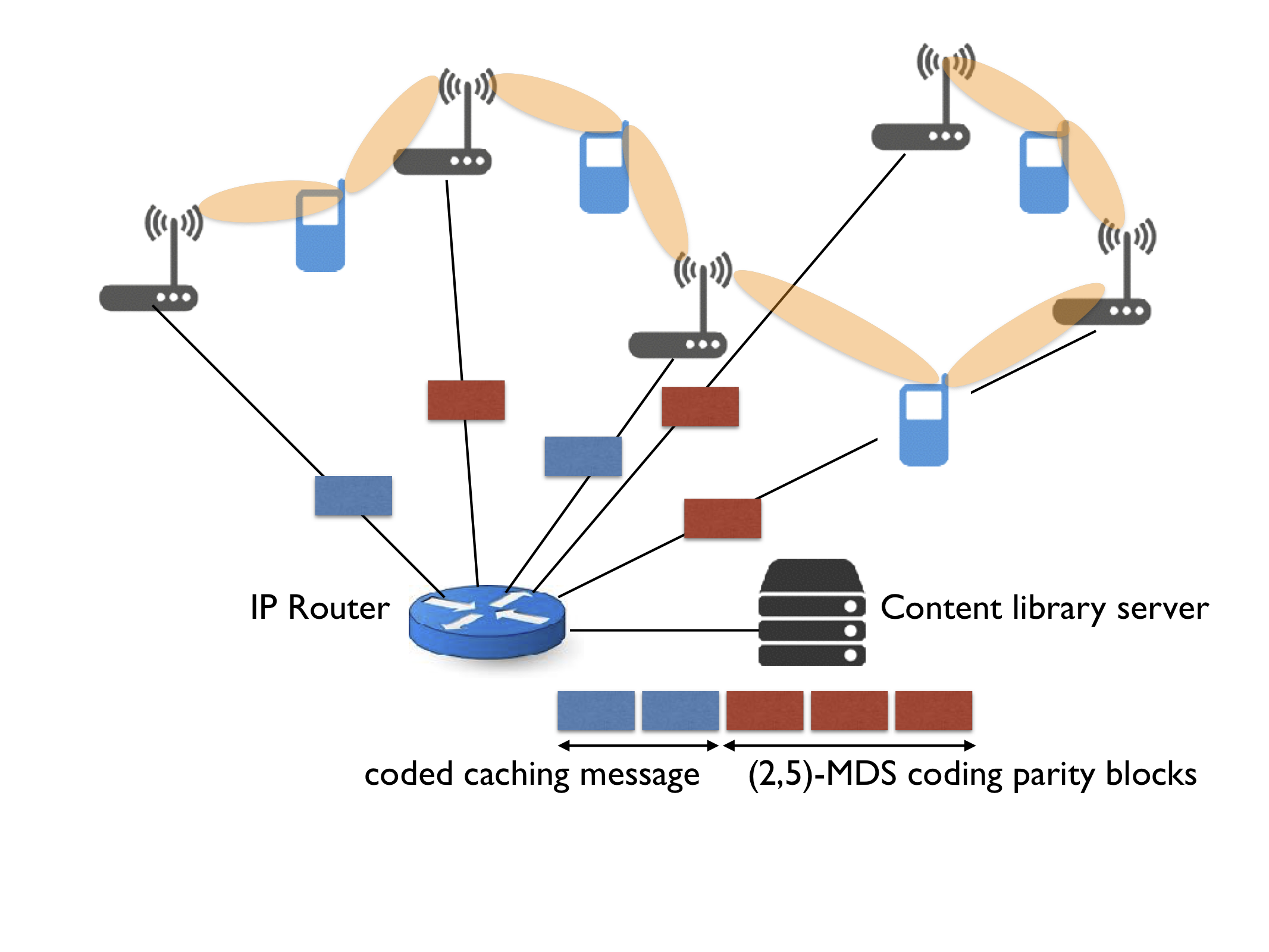}
	\caption{\small A conceptual example of the proposed system with $N_E = 5$ and $L = 2$. User nodes make use of their beamforming antenna array to
	decode two out of five EN transmissions, by treating the others as noise.}
	\label{system-scheme}
\end{figure}

The users make use of their antenna array to detect and decode the PHY codewords of $L$ ENs. Thanks to MDS coding, whenever a user $k$
is able to decode $L$ ENs messages, then it can reconstruct the whole multicast codeword $X$. In fact, the MDS condition is that the information 
message $X$ can be reconstructed from any $L$ out of $N_E$ MDS-coded blocks. 
At this point, user $k$ can decode the desired file $d_k$ from $X$ and its cache content, as detailed in \cite{maddah2014fundamental}.  
Fig.~\ref{system-scheme} shows an example with $N_E = 5$ and $L = 2$, where all users that can decode 2 data streams 
from two distinct ENs can retrieve their requested file.

The delivery latency, i.e., the time necessary to deliver all requests, is given by 
\begin{equation}  \label{delivery-latency}
D  = \frac{1}{LR} \times \frac{F}{w} \times \frac{K (1 - \mu)}{1 + K \mu},  
\end{equation}
where $w$ is the bandwidth of the wireless channel and  $R$ is the PHY rate, also referred to as {\em spectral efficiency}, expressed in bit/s/Hz. 
Notice that $D$ in (\ref{delivery-latency}) is given by the product 
of three factors: $F/w$ depends on the file size and the system bandwidth (baudrate), and is fixed by the system design. 
The term $\frac{K (1 - \mu)}{1 + K \mu}$ is the coded caching gain as in \cite{maddah2014fundamental}. 
Finally, $1/(LR)$ is the combined effect of MDS ``macro-diversity order'' and the PHY coding rate. {\em Our main system design goal consists of
determining $L$ such that the product $LR$ is maximized (i.e., the delivery latency is minimized) 
for a given target per-user outage probability.} 

From the PHY viewpoint, the goal for each user is  to successfully decode the $L$ strongest ENs. 
Assuming that the PHY codewords are long enough such that the small-scale Rayleigh fading behaves as a stationary 
and ergodic process over the time-frequency slot spanned by a PHY codeword,  
we denote by ${C}_{k,\ell}(\Phi)$ the achievable ergodic rate that user $k$ can successfully decode {\em with high probability} (w.h.p.) from EN $\ell$
for a given decoding strategy (to be detailed later) and a given geometry $\Phi$, i.e., a given placement of the ENs relative to user $k$.
Without loss of generality, we can sort these achievable rates such that ${C}_{k,1}(\Phi), \ldots, {C}_{k,L}(\Phi)$ are the largest rates
over all the NEs.  It follows that user $k$ can obtain its requested file if 
\begin{equation} \label{decoding-condition}
{C}_{k}^{(L)}(\Phi) = \min \{ {C}_{k,\ell}(\Phi) : \ell \in [L] \} > R,
\end{equation} 
For a random geometry $\Phi$, we can define the user outage probability as 
\begin{equation} \label{outagep} 
P_{{\rm out},k}(R) = \PP( \min_{\ell \in [L]} {C}_{k,\ell}(\Phi) \leq R), 
\end{equation}
given by the  Cumulative Distribution Function (CDF) of $\min_{\ell \in [L]} {C}_{k,\ell}(\Phi)$.

\section{System performance analysis}
\label{sec:StatementOfProblem}

We  assume that ENs are randomly located on the two-dimensional plane 
and are distributed according to a two-dimensional homogeneous Poisson Point Process (PPP) of density $\lambda$.
The ensemble of the locations of the ENs is denoted by $\Phi$. 
The propagation is characterized by small-scale Rayleigh fading that remains constant on time-frequency coherence blocks of $n$ symbols 
(block fading model) and evolves according to a stationary ergodic process from block to block. 
Each receiving user acquires the downlink (DL) channel state information at the receiver (CSIR) from its surrounding ENs (e.g., through common DL pilots periodically broadcasted by the ENs). For simplicity, we do not consider CSIR error and pilot overhead in the present analysis. However, broadcasting downlink pilots, or ``beacon'' signals, from the base stations/access points is common practice of any current cellular system and WiFi network, so that the same type of overhead and (small) performance degradation incurred by any standard technology is to be expected in our system.

Given the symmetry of the problem, without loss of generality we can consider the performance of the typical user indexed by $k$ and located at the origin. We sort the ENs according to their distance from user $k$ 
in non-decreasing order, and let $j\in [N_E]$ be the index of the $j$-th closest EN to user $k$.  
The space-time signal received by user $k$ is given by 
\begin{equation}
\Ym_k = \sum_{j=1}^{N_E}  \sqrt{\beta r_{k,j}^{-\eta}}   \hv_{k,j} \underline{\xv}_j + \Nm_k
\label{eq:rec_sig}
\end{equation}
where $\Ym_k \in \CC^{n_r \times n}$ with $n$ being the size of the fading blocks in symbols, 
$\hv_{k,j} \in \CC^{n_r \times 1}$  is the channel vector containing the small-scale fading coefficients from EN $j$ to the array of user $k$, 
$\underline{\xv}_j \in \CC^{1 \times n}$ is the coded-modulation block of symbol sent by EN $j$, 
$r_{k,j}$ is the distance between EN $j$ and user $k$, $\eta$ is the pathloss exponent, 
and $\beta$ is the path loss intercept.  The ENs transmit at a constant average power $\frac{1}{n} \EE[ \underline{\xv}_j  \underline{\xv}^\herm_j ] = P$.
The noise samples in the matrix $\Nm_k$ are independent and identically distributed (IID) $\sim \Cc\Nc(0,N_0)$. 
According to the independent Rayleigh statistics,  ${\hv}_{k,j}$ are IID across users and ENs and have IID components 
$\sim \mathcal{CN}(0,1)$.

We introduce here the \textit{Partial Zero-Forcing} (PZF) receiver strategy, proposed for our system.  
Let $\Hm_k = [\hv_{k,1}, \ldots, \hv_{k,N_E}]$ denote the $n_r \times N_E$ matrix of the channel coefficients from the EN antennas to user $k$ antenna array, 
and, for a subset $\Bc \subseteq [N_E]$, let $\Hm_{k, \Bc}$ denote the submatrix formed by the columns with indices $j \in \Bc$. 
PZF consists of applying linear zero-forcing only with respect to the signals of the $L$ nearest ENs, 
whereas the signals of the remaining ENs with indices  $j > L$  are treated as noise.  
Assuming $n_r \geq L$, the receive filtering matrix of user $k$ is defined as the pseudo-inverse of the $n_r \times L$ 
matrix $\Hm_{k,1:L}$. This is given by 
\begin{equation}
\Hm_{k,1:L}\ssymbol{2} = \Hm_{k,1:L} ( \Hm^\herm_{k,1:L}\Hm_{k,1:L})^{-1}. \label{eq:psuedo_invers}
\end{equation}
Denoting by $\hv_{k,\ell}\ssymbol{2}, 1 \leq \ell \leq L$, the $\ell$-th column of $\Hm_{k,1:L}\ssymbol{2}$ and by $\qv_{k,\ell} = \hv_{k,\ell}\ssymbol{2} / \lVert \hv_{k,\ell}\ssymbol{2} \rVert$ the normalized receive filtering vector corresponding to the $\ell$-th stream, the filtered output is written as
\begin{align}
\tilde{\underline{\yv}}_{k,\ell} &= \qv_{k,\ell}^\herm \Ym_{k} \nonumber \\
&= \frac{(\hv_{k,\ell}\ssymbol{2})^\herm}{ \lVert \hv_{k,\ell}\ssymbol{2} \rVert } \hv_{k,\ell} \sqrt{\beta r_{k,\ell}^{-\eta}} \, \underline{\xv}_{\ell}  +  \sum_{j= L+1}^{N_E} \frac{(\hv_{k,\ell}\ssymbol{2})^\herm}{ \lVert \hv_{k,\ell}\ssymbol{2} \rVert }  \hv_{k,j} \sqrt{\beta r_{k,\ell}^{-\eta}} \, \underline{\xv}_{j}  + \underline{\tilde{\zv}}_k  \label{userful}
\end{align}
where $\underline{\tilde{\zv}}_k$ is the filtered noise vector with IID components $\sim \Cc\Nc(0,N_0)$. 

In the rest of this paper, for the sake of analytical tractability, we focus on the case of an infinitely extended network (i.e., $N_E \rightarrow \infty$) and high-SNR interference limited performance (i.e., $N_0 \rightarrow 0$).  As a result, the system becomes ``scale-free'', that is, the results of our analysis 
are invariant to any non-zero scaling factor of the physical distances $r_{k,\ell}$ appearing in  (\ref{userful}).

The Signal-to-Interference Ratio (SIR) at user $k$ receiver is given by 
\begin{eqnarray} \label{sir-given-phi}
\SIR_{k,\ell} &=& \frac{ r_{k,\ell}^{-\eta}  \lVert \hv_{k,\ell}\ssymbol{2} \rVert^{-2}}{\sum_{j= L+1}^{\infty} r_{k,j}^{-\eta}   | \tilde{h}_{k,j} |^2 }. 
\end{eqnarray}
Since $\qv_{k,l}$ is a unit vector that is independent of $\{ \hv_{k,j} : j  = L+1, L+2, \dots \}$, it follows that $\tilde{h}_{k,j} = \qv_{k,l}^\herm \hv_{k,j} \sim \mathcal{CN}(0,1), \forall j \geq L+1$. Furthermore, a well-known property of the ZF linear detector is that in the case of Gaussian IID channel 
vectors the useful signal coefficient
$\lVert \hv_{k,l}\ssymbol{2} \rVert^{-2} = \mathcal{X}_{2(n_r-L+1)}$ is Chi-squared distributed with $2(n_r - L+1)$ degrees of 
freedom \cite{massiveMIMO-book-Marzetta} and mean $\EE[\mathcal{X}_{2(n_r-L+1)}] = n_r - L + 1$. 
The the ergodic achievable rate with Gaussian inputs and treating interference as noise 
for the transmission of the $\ell$-th closest EN at user $k$ is given by 
\begin{equation} \label{cazzo1}
{C}_{k,\ell}(\Phi) = \EE \left [ \left . \log \left (1+    \frac{ r_{k,\ell}^{-\eta} \mathcal{X}_{2(n_r-L+1)} }{\sum_{j= L+1}^{\infty} r_{k,j}^{-\eta}  | \tilde{h}_{k,j} |^2 } \right )  \right  |\Phi \right ],  
\end{equation}
where the random variables $\mathcal{X}_{2(n_r-L+1)}$ and $ | \tilde{h}_{k,j} |^2 : j > L$ are mutually independent, and the conditioning 
with respect to the geometry $\Phi$ determines the distances $r_{k,\ell}$ and $r_{k,j} : j > L$. 
In order to obtain analytically tractable expressions, we first apply Jensen's inequality with respect to the interference fading coefficients
$| \tilde{h}_{k,j} |^2$, $j \geq L+1$. Using $\EE[| \tilde{h}_{k,j} |^2] = 1$ and the convexity of the function $f(x) = \log(1 + a/x)$ for $a > 0$ and $x > 0$, 
we obtain the achievable rate lower bound
\begin{align} \label{cazzo2}
{C}_{k,\ell}(\Phi) 
&=  \EE \left [ \left . \log \left (1 +  \rho_{k,\ell} \mathcal{X}_{2(n_r-L+1)}  \right )  \right  |\Phi \right ].  
\end{align}
where we define the {\em local-average SIR} 
\begin{equation} \label{eq:actual_SIR}
{\rho}_{k,\ell} = \frac{r_{k,\ell}}{\sum_{j= L+1}^{\infty} r_{k,j}^{-\eta}}, \quad 1\leq \ell \leq L.
\end{equation}
that captures the network geometry.
Next, we would like to eliminate the dependency on the distances $r_{k,j}$ with $j \geq L+1$. To this purpose, we replace the conditioning with respect to $\Phi$ by the conditioning with respect to only $r_{k,\ell}$ and $r_{k,L}$, i.e., the $\ell$-th and the $L$-th shortest distances from the origin. 
Furthermore, letting $I = \sum_{j= L+1}^{\infty} r_{k,j}^{-\eta}$ and noticing that $r_{k,j}$ for $j \geq L+1$ depends on 
PPP points outside the disk of radius $r_{k,L}$ around user $k$ while $r_{k,\ell}$ depend on PPP points inside the disk, we have that \cite{haenggi2012stochastic}
$\EE[ I | r_{k,L}, r_{k,\ell} ] = \EE[ I | r_{k,L} ]$.
By applying this approximation and Jensen's inequality in (\ref{cazzo2}), we obtain
\begin{align}
{C}_{k,\ell}(\Phi)  &\gtrapprox 
\EE \left [ \left . \log \left (1 + \frac{ r_{k,\ell}^{-\eta} \mathcal{X}_{2(n_r-L+1)} }{\sum_{j= L+1}^{\infty} r_{k,j}^{-\eta}} \right )  \right  | r_{k,L} , r_{k,\ell} \right ] \label{cazzo4} \\
& \geq    
\EE \left [ \left . \log \left (1 + \frac{ r_{k,\ell}^{-\eta} \mathcal{X}_{2(n_r-L+1)} }{\EE\left [ \left . I \right | r_{k,L} \right ]} \right )  \right  | r_{k,L} , r_{k,\ell} \right ] .  \label{cazzo3}
\end{align}
In order to calculate the conditional average interference term in (\ref{cazzo3}), 
we apply Campbell theorem \cite[Theorem 4.1]{haenggi2012stochastic}  and, for $\eta > 2$, we obtain
\begin{eqnarray}
\mathbb{E}[I | r_{k,L}] = \frac{2 \pi \lambda}{\eta -2} r_{k,L}^{2-\eta}.
\label{lemma: interference}
\end{eqnarray}
Plugging (\ref{lemma: interference}) in (\ref{cazzo3}), we obtain the quasi-lower bound\footnote{This is referred to as quasi-lower bound because of the approximation in \eqref{cazzo4}.} on the ergodic achievable rate in the form
\begin{equation} \label{Clb}
{C}_{k,\ell}^{\rm qlb}(\Phi) = \EE \left[ \left . \log \left(  1+  \tilde{ \rho}_{k,\ell} \mathcal{X}_{2(n_r-L+1)} \right ) \right | \Phi \right].
\end{equation}
where we define the approximated conditional {local-average SIR} as 
\begin{align}  \label{ziopippa}
\tilde{\rho}_{k,\ell} & =  \frac{r_{k,L}^{\eta-2}}{r_{k,\ell}^{\eta} } \frac{\eta-2}{2 \pi \lambda}  \quad 1\leq \ell \leq L .
\end{align}
Notice that $\tilde{\rho}_{k,\ell}$ is related to $\rho_{k,\ell}$ defined in \eqref{eq:actual_SIR} by replacing the denominator $I$ of  \eqref{eq:actual_SIR} 
with its conditional expectation given in \eqref{lemma: interference}. 
This approach yields analytically tractable expressions and 
is expected to provide a very accurate approximation of the actual achievable ergodic rate for fixed geometry $\Phi$ because the interference term $I$ 
contains the cumulative effect of many ``far'' ENs, such that it is expected that a sort of self-averaging effect kicks in. 

\begin{figure}
\centering
\includegraphics[scale=0.25]{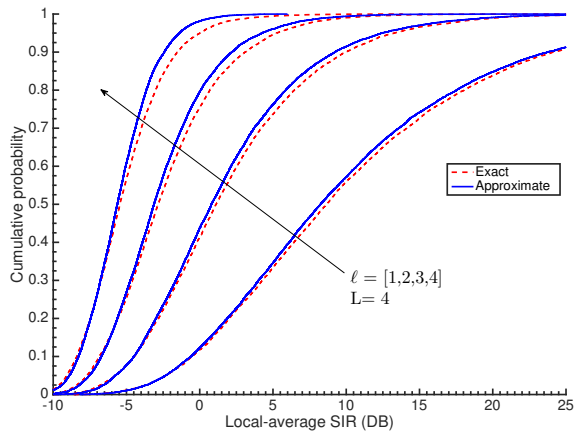}
\caption{\small CDFs of the exact and approximate local-average SIRs ${\rho}_{k,\ell}$ and $\tilde{\rho}_{k,\ell}$ with $n_r=8$. }
\label{fig:ordering}
\end{figure} 

Fig.~\ref{fig:ordering} compares the CDF of the local average SIR $\tilde{\rho}_{k,\ell}$ obtained via approximation 
in (\ref{ziopippa}) against its exact form ${\rho}_{k,\ell}$ in (\ref{eq:actual_SIR}). All the curves are obtained via Monte-Carlo simulations by setting the EN density $\lambda= 8 \, \mathrm{ENs/km}^2$ (which amounts to on average one EN in a circular cell of radius $200$ m), the pathloss exponent $\eta=3.75$. We assumed that the PZF is applied to cancel the $L=4$ nearest interferers at the receiving user equipped with antenna array size $n_r=8$.
As it can be seen from Fig.~\ref{fig:ordering}, the CDFs of $\tilde{\rho}_{k,\ell}$ and  ${\rho}_{k,\ell}$ are very close for 
a wide range of SIR. Similar results have been observed for a great variety of system parameters. 

\subsection{Main Results}
\label{sec:MainResults}

We first present some distance distributions, which shall be useful in subsequent derivations.
Given a homogeneous PPP $\Phi$ of density of $\lambda$, let $r_n$ denote the 
$n$-th shortest distance of points of $\Phi$ from the origin.
Then, the Probability Density Function (PDF) of $r_n$ is given by~\cite{distance-distrib-haenggi}
\begin{equation}
f_{ r_{n}}(v) = \frac{2(\pi \lambda)^{n}}{(n-1)! } v^{2n-1} e^{-\pi \lambda v^2},  \;\; v \geq 0, \label{eq:marginal_pdf}
\end{equation}
and the joint PDF of $r_{\ell}$ and $r_{n}$ with $1 \leq \ell < n$ is given by \cite{martin2014multi}
\begin{align}
f_{r_{\ell}, r_{n}}(u,v) &= \frac{4 (\pi \lambda)^{n}}{(n-\ell-1)! \, (\ell - 1)!}(v^2-u^2)^{n-\ell-1} v u^{2\ell-1} e^{-\pi \lambda v^2}, \;\;\; u,v \; \geq 0.
 \label{eq:jointrr}
\end{align}
Using these results, we next provide a closed-form accurate approximations for the outage probability and the average spectral efficiency 
in terms of the macro-diversity order $L$, the receive antenna array size $n_r$, the pathloss exponent $\eta$, and the delivery rate $R$.

\subsubsection{Outage probability}  \label{outage-section}

Based on the spatial distribution of the EN locations we can derive a distribution for the local-average SIR. Due to the difficulty of obtaining an exact 
closed-form for this distribution, we derive an approximate expressions based on the approach introduced in \cite{blaszczyszyn2016spatial, abate1995numerical}. First, the Laplace transform of the CDF of $1/\tilde{\rho}_{k,\ell}$ is derived. Then, this is numerically inverted via Euler series expansion. 
Finally, the CDF can be obtained in the form of finite series yielding accurate results with only a few terms.
Specifically, from \cite{blaszczyszyn2016spatial}, we obtain the sought approximated form for the CDF of $1/\tilde{\rho}_{k,\ell}$ as
\begin{align}  \label{rhocdf}
F_{\tilde{\rho}_{k,\ell}} (\gamma) &\approx  1- \gamma \frac{e^{A/2}}{2^B}\sum_{b=0}^{B}  {{B} \choose {b}} \sum_{g = 0}^{G+b}\frac{(-1)^g}{D_g} {\rm Re} \left \{\frac{\mathcal{L}_{1 /\tilde{\rho}_{k,\ell}(\tau)}}{\tau}\right \},
\end{align}
where 
\begin{equation}
\tau =  \frac{(A + \mathrm{i} 2\pi g) \gamma}{2} .
\end{equation}
The values $D_0 =2$ and $D_g=1$ for $g>0$ and  $A=9.21$, $B= 5$ and $G= 8$ provide a satisfactory accuracy \cite{o1997euler}. 
The Laplace transform of $1/\tilde{\rho}_{k,\ell}$, for $s \in \CC$, is given by 
\begin{align}  
\mathcal{L}_{1 /\tilde{\rho}_{k,\ell}}(s) &= \EE \left [e^{-s /\tilde{\rho}_{k,\ell}} \right ]  \\
&=  \int_{0}^{\infty} \Dd v \int_{0}^{v}  e^{-s \frac{u^{\eta} }{v^{\eta-2}} \frac{2 \pi \lambda}{\eta-2}} f_{r_{k,\ell}, r_{k,L}}(u, v) \Dd u \label{ziocanale}
\end{align}
where $f_{r_{k,\ell}, r_{k,L}}(u, v)$ is given by (\ref{eq:jointrr}) with the replacement $n \leftarrow L$.  
Evaluating the integral in (\ref{ziocanale}) separately for the two cases $\ell < L$ and $\ell= L$ yield the following compact expressions, proved in 
Appendix~\ref{ap:Laplace_interference}. 

\textbf{Case 1:} For $\ell<L$ 
\begin{align}
	\mathcal{L}_{1 /\tilde{\rho}_{k,\ell}}(s) = &\sum_{n= 0}^{L-\ell-1} \frac{ (-1)^n \Gamma(L)}{(L-\ell-1)\, ! \, (\ell-1)\, !  \, (n+\ell)} {{L-\ell-1} \choose {n}}   \Hypergeometric{2}{1}{L,\eta'}{\eta'+1}{\frac{-2s}{\eta-2}},
	\label{eq:Laplace_ell}
\end{align}
	where $\eta'= 2\frac{n+\ell}{\eta}$ and $   \Hypergeometric{2}{1}{a,b}{c}{z}$	is the Gaussian hypergeometric function.
	
\textbf{Case 2:} For $\ell= L$ 
	\begin{eqnarray}
	\mathcal{L}_{1 /\tilde{\rho}_{k,\ell}}(s) = \left( \frac{2s}{\eta-2} + 1\right)^{-L} .
	\label{eq:Laplace_L}
	\end{eqnarray}
	

\begin{figure}[t]
	\centering
	\includegraphics[scale=0.25]{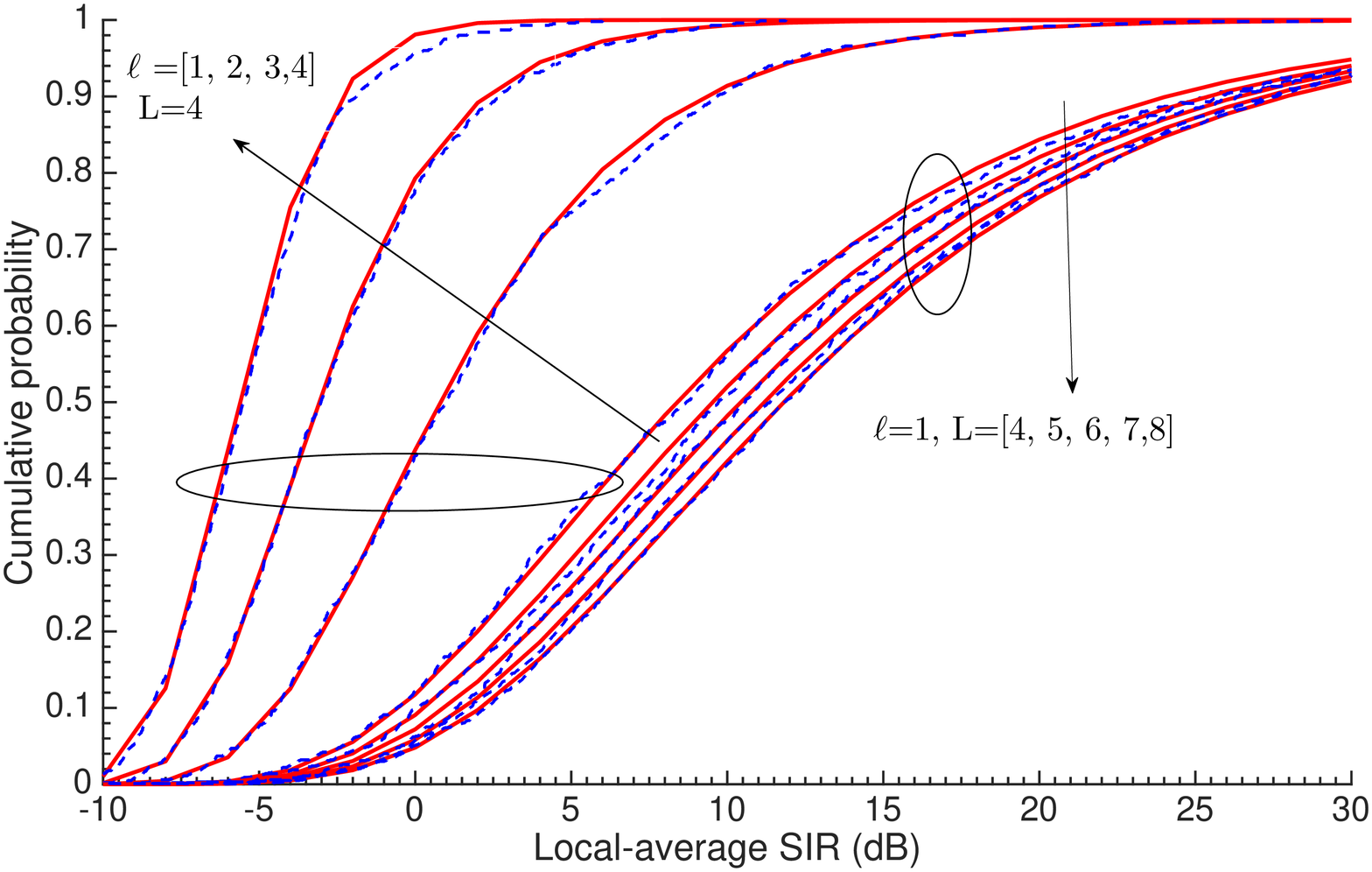}
	\caption{\small CDF of the approximated local-average SIR ($\tilde{\rho}_{k,\ell}$ )
	with $n_r=8$. Solid curves correspond to the analysis in (\ref{rhocdf}) while the dashed curves are obtained via simulation.}
	\label{fig:local_sir}
\end{figure}

Fig. \ref{fig:local_sir} illustrates the CDF of the approximated local-average $\SIR$ given in (\ref{rhocdf}) with different sub-stream indices 
and different values of $L$ with a receive antenna array size $n_r=8$, EN density $\lambda= 8 \, \mathrm{ENs/km}^2$, 
and $\eta=3.75$. Monte Carlo simulation of the CDF is also reported for comparison, and shows that our 
analytical approximation is very accurate.  

The expectation with respect to $\mathcal{X}_{2(n_r-L+1)}$ in (\ref{Clb})
can be calculated in closed form as
\begin{eqnarray}  \label{Clb-explicit}
{C}_{k,\ell}^{\rm qlb}(\tilde{\rho}_{k,\ell})
&= &\EE \left[ \left . \log \left( 1+  \tilde{\rho} _{k,\ell} \mathcal{X}_{n_r-L+1}\right) \right | \Phi \right] \nonumber\\
&=& \mathcal{I}_{(n_r-L+1)} ( \tilde{\rho}  _{k,\ell})\log (e),
\end{eqnarray}
where with a slight abuse of notation we use ${C}_{k,\ell}^{\rm qlb}(\tilde{\rho} _{k,\ell})$ instead of 
${C}_{k,\ell}^{\rm qlb}(\Phi)$ since the dependency on the PPP points reduces to the single random variable $\tilde{\rho}_{k,\ell}$, and where
we define the expression (see \cite{alouini1999capacity})
\begin{align}
\label{eq:exp_int}
\mathcal{I}_M(\mu) ~= ~&\EE \left[ \log_e (1+\mu\mathcal{X}_{2M})\right]  \nonumber \\
= ~&  \Pi_M(-1/\mu)E_i(1,1/\mu) + \sum_{m=1}^{M-1}\frac{1}{m} \Pi_m(1/\mu) \Pi_{M-m}(-1/\mu),
\end{align}
with $ \Pi_n(x) = e^{-x}\sum_{i=0}^{n-1}\frac{x^i}{i!}$ and with the exponential integral function defined as
$E_i(n,x) = \int_1^\infty t^{-n} e^{-xt} dt$.
Due to the monotonicity of the log function, ${C}_{k,\ell}^{\rm qlb}(\tilde{\rho} _{k,\ell})$ is strictly monotonically increasing in $\tilde{\rho} _{k,\ell}$.
Then, using ${C}_{k,L}^{\rm qlb}(\tilde{\rho} _{k,\ell})$ {\em instead} of $ {C}_k^{(L)}(\Phi)$
in the outage probability expression (\ref{outagep}), we arrive at the desired quasi-upper bound to the outage probability in the form
\begin{eqnarray} \label{outagep1} 
P^{\rm out}_{k}(R) &  \lessapprox & \PP\left ( {C}_{k,L}^{\rm qlb}(\tilde{\rho} _{k,L}) \leq R \right ) \nonumber \\
& = & \PP \left ( \tilde{\rho} _{k,L} \leq \left ( {C}_{k,L}^{\rm qlb}\right )^{-1} (R) \right )  \nonumber \\
& = & F_{\tilde{\rho} _{k,L}} \left ( \left ( {C}_{k,L}^{\rm qlb}\right )^{-1} (R)  \right ),
\end{eqnarray}
where the inverse function $\left ({C}_{k,L}^{\rm qlb}\right )^{-1}(\cdot)$ is easily obtained numerically from (\ref{Clb-explicit}). 

\subsubsection{Average spectral efficiency}

In order to appreciate the tradeoff between the macro-diversity order $L$ and the achievable physical-layer transmission rate $R$, 
and to quickly see which value of $L$ minimizes the delivery latency, we can consider the average spectral efficiency of a typical user
when averaging is also with respect to the random locations of the ENs (i.e., with respect to $\Phi$). We indicate by ``overline'' 
quantities averaged also with respect to  the geometry. Then, we have
\begin{align} 
\overline{C}_{k,\ell} &= \EE \left [  \log \left (1 + \frac{ r_{k,\ell}^{-\eta} \mathcal{X}_{2(n_r-L+1)} }{\sum_{j= L+1}^{\infty} r_{k,j}^{-\eta}} \right )   \right ] \\
&=\EE \left [ \EE \left [ \left . \log \left (1 + \frac{ r_{k,\ell}^{-\eta} \mathcal{X}_{2(n_r-L+1)} }{\sum_{j= L+1}^{\infty} r_{k,j}^{-\eta}} \right )  \right  | r_{k,L} , r_{k,\ell} \right ] \right]\\
&\geq \EE \left[ \EE \left [ \left . \log \left (1 +  \tilde{\rho} _{k,\ell} \mathcal{X}_{2(n_r-L+1)}  \right )  \right  | r_{k,L} , r_{k,\ell} \right ]  \right]\\
&= \EE \left[ \log_2 \left(  1+  \tilde{\rho} _{k,\ell} \mathcal{X}_{2(n_r-L+1)} \right )\right]  \label{lowerbound} 
\end{align}

\begin{remark} 
Comparing (\ref{lowerbound}) with (\ref{Clb}) and using the law of iterated expectation, for which 
$\EE\left [\log \left(  1+  \tilde{ \rho}_{k,\ell} \mathcal{X}_{2(n_r-L+1)} \right ) \right ] = \EE\left [ \EE \left[ \left . \log \left(  1+  \tilde{ \rho}_{k,\ell} \mathcal{X}_{2(n_r-L+1)} \right ) \right | \Phi \right] \right]$, we have that the expectation of our quasi-lower bound ${C}_{k,\ell}^{\rm qlb}(\Phi)$
on the ergodic mutual information for a given geometry $\Phi$ yields a true lower bound $\overline{C}_{k,\ell}^{\rm lb}$
on the  average (w.r.t. the random geometry) ergodic mutual information. 
	\hfill $\lozenge$
\end{remark}
At this point, it is convenient to define 
\begin{equation}
\widetilde{\SIR}_{k,\ell} = \tilde{\rho} _{k,\ell} \mathcal{X}_{2(n_r-L+1)},
\end{equation}
as the ``SIR'' quantity appearing in (\ref{lowerbound}) and evaluate the expectation by following the approach of \cite{MunMorLoz-TWC15}. We can write
\begin{align}
\overline{C}_{k,\ell}^{\rm lb} 
&= \int_{0}^{\infty} \log_2(1+\gamma) \, \Dd F_{\widetilde{\SIR}_{k,\ell}}(\gamma) \\
&= \int_{0}^{\infty} \left(1-F_{\widetilde{\SIR}_{k,\ell}}(2^{\gamma}-1)\right) \, \Dd \gamma . \label{eq: avg_spec}
\end{align}

Notice that, for fixed $\tilde{\rho} _{k,\ell}$, $\widetilde{\SIR}_{k,\ell}$ is chi-squared distributed (up to a scaling factor). Hence,
its conditional CDF is given by
\begin{equation}
F_{\widetilde{\SIR}_{k,\ell} | \tilde{\rho} _{k,\ell} } (\gamma)  =  1-\frac{\Gamma \left(n_r-L+1,   \gamma/\tilde{\rho} _{k,\ell} \right )}{\Gamma{(n_r-L+1)}}.
\label{eq:chi_sir}
\end{equation}
Where $\Gamma(a,x) = \int_{x}^{\infty}  t^{a-1} e^{-t}dt$ is  the upper  imcoplete gamma function.
The unconditional CDF of $\widetilde{\SIR}_{k,\ell}$ is derived by expressing $\tilde{\rho} _{k,\ell}$ in terms of $r_{k,\ell}$ and $r_{k,L}$ (cf. (\ref{ziopippa})) and using their joint PDF to average these variable out. The following expressions for $F_{\widetilde{\SIR}_{k,\ell} } (\cdot)$ are obtained by treating the two cases $\ell < L$ and $\ell = L$ separately (cf. Appendix \ref{ap:proof_prop2}).
	
\textbf{Case 1:} For $\ell < L$ 
\begin{align}
F_{\widetilde{\SIR}_{k,\ell}}(\gamma) &= 1 - \sum_{ m= 0}^{n_r-L}\sum_{n= 0}^{L-\ell-1}  \frac{2 (-1)^n }{m\, ! \, (L-\ell-1)\, ! \, (\ell-1)\, !  } \frac{\Gamma(m+L)}{\eta \, (\eta'+m)} {{L-\ell-1} \choose {n}} \nonumber \\
&\qquad \quad \cdot  \left(\frac{2 \gamma}{\eta-2} \right)^m  \Hypergeometric{2}{1}{L+m,m+\eta'}{m+\eta'+1}{-\frac{2 \gamma}{\eta-2}}  \label{eq:CDF_SIR_ell} 
\end{align}
	
\textbf{Case 2:} For $\ell= L$ 
\begin{equation} \label{eq:CDF_SIR_L}
F_{\widetilde{\SIR}_{k, \ell}}(\gamma) =1- \sum_{ m= 0}^{n_r-L} \frac{(m+L-1)!}{m\, !\, (L-1)\,!\,} \frac{(\frac{2\gamma}{\eta-2} )^m}{(\frac{2\gamma}{\eta-2}+1)^{(m+L)}} .  
\end{equation}
Finally, by invoking (\ref{eq:CDF_SIR_ell}) and (\ref{eq:CDF_SIR_L}) in (\ref{eq: avg_spec}) and evaluating the integrals, the lower bound of the average ergodic spectral efficiency corresponding to the $\ell$-th data stream can be expressed as follows (cf. Appendix~\ref{ap:proof_prop3}):
	
\textbf{Case 1: } For $\ell<L$ 
\begin{align}
\overline{C}_{k,\ell}^{\rm lb}  &= \sum_{ m= 0}^{n_r-L}\sum_{n= 0}^{L-\ell-1}  \frac{2 (-1)^n }{m\,!\,(L-\ell-1)\,!\, (\ell-1)\,!\, \eta } {{L-\ell-1} \choose {n}}   \nonumber \\
&\quad \cdot  G_{3,3}^{2,3} \left(\begin{matrix} -(m+1), -(m+L),-m-\eta' \\
-(m+1),  -1,-(m+1+\eta')
\end{matrix} \bigg| \frac{2}{\eta-2} \right)  \left(\frac{2}{\eta-2}\right)^{m+1}   \log_2(e) \label{eq:Avg Seff ell expr}         
\end{align}
where $G_{m,n}^{p,q} \left(\begin{matrix} a_1,\dots, a_n,a_{n+1},\dots, a_p \\
b_1, \dots, b_m, b_{m+1}, \dots, b_q 
\end{matrix} \bigg| z\right)  $ is Meijer-G function.

\textbf{Case 2: } For $\ell= L$ 
\begin{align}
\label{eq: sp_worst}
\overline{C}_{k,L}^{\rm lb}  &= \sum_{ m= 0}^{n_r-L} \frac{\log_2(e) }{m+L} \Hypergeometric{2}{1}{1,L}{m+L+1}{1-\frac{2}{\eta-2}}.
\end{align}

\begin{figure}[t]
	\centering
	\includegraphics[scale=0.25]{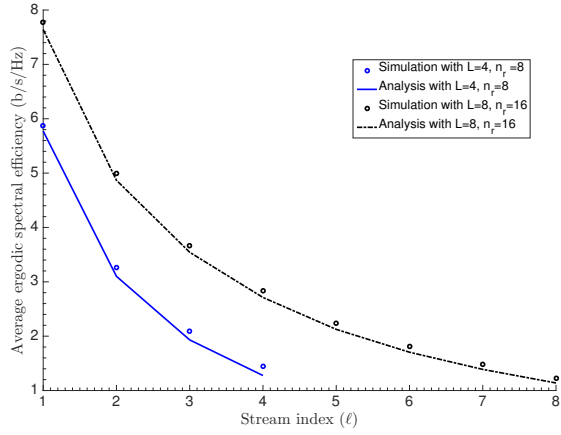}
	\caption{\small Average ergodic spectral efficiency versus stream index.}
	\label{fig:spectral}
\end{figure}

The tightness of this lower bounds is illustrated in Fig. \ref{fig:spectral}, comparing 
$\overline{C}_{k,L}^{\rm lb}$  and $ \overline{C}_{k,L} $ for the 
settings: $\lambda= 8 \, \mathrm{ENs/km}^2$ and $\eta = 3.75$ with $n_r = 8, L=4$ and with $n_r=16, L=8$.
Since eventually the spectral efficiency is defined by the worst-case data stream, we define the lower bound for minimum average ergodic spectral efficiency as follows
\begin{equation}
\overline{C}_{k,{\rm PZF}}^{\rm lb} = \overline{C}_{k,L}^{\rm lb}.
 \label{eq:lb_pzf}
\end{equation}

\begin{remark} {\bf Operational significance of the average ergodic spectral efficiency.}
Since each user must receive $L$ data streams from $L$ nearest ENs, assuming that the typical user is able to receive at common rate 
$R = \overline{C}_{k,{\rm PZF}}^{\rm lb}$ bit.s.Hz, 
we can consider the value of $L$ that maximizes the product $L \times \overline{C}_{k,{\rm PZF}}^{\rm lb}$  and yields the minimum delivery time. 
This has an operational significance if the average ergodic rate was indeed achievable. 
In turns, this depends on how quickly the geometry changes with respect to the duration of a codeword. 
We remark here that in classical stochastic geometry analysis it is customary to mix the time dynamics of the small-scale fading and of 
the random geometry. As a matter of fact, the small-scale fading ``mixes'' (i.e., it loses memory) on time intervals 
of the order of the inverse of the Doppler bandwidth, i.e., typically between 1 and 10 ms. 
In contrast, the geometry (distances between a user and the ENs) evolves according to much slower dynamics, 
and typically can be considered constant on intervals between 1 and 10 s for users moving at vehicular speeds, and even much larger 
for nomadic users.  Hence, the operational meaning of the average ergodic rate computed here in terms of achievable rate 
is questionable, and this is precisely the reason for which we focused mainly on the outage probability, 
where outages are defined with respect to the random  geometry.  Nevertheless, the product $L \times \overline{C}_{k,{\rm PZF}}^{\rm lb}$ 
captures  (in an average sense) the tradeoff between $L$ and $R$.  For a desired value of this product, making the EN messages shorter by 
increasing $L$ allows to reduce the transmission rate $R$, but the users are required to decode signals from more ENs, such that eventually the outage 
probability of the worst stream (the $L$-th stream) becomes too large. 
\hfill $\lozenge$
\end{remark}
\subsubsection{ PZF with Successive Interference Cancellation}

Previously we considered the linear PZF detection strategy at the users 
for its analytical tractability. In this section we improve the PZF strategy by introducing successive interference cancellation (SIC).  
In PZF-SIC users decode their strongest EN signals in sequence. 
We let $\Pi$ denote the set of permutations of $[L]$, and let  $\pi = \{ \pi(1), \pi(2), \dots, \pi(L) \} \in \Pi$ denote the decoding order 
of a generic user $k$. Then,  user $k$ decodes $\underline{\xv}_{\pi(1)}$ first, 
by suppressing interference from the remaining $L-1$ ENs via ZF beamforming, 
Then, it subtracts the decoded version of $\underline{\xv}_{\pi(1)}$ from the received signal, and proceeds to the decoding of the next signal 
$\underline{\xv}_{\pi(2)}$ by suppressing interference from the remaining $L-2$ ENs via ZF beamforming. 
Notice that since we treat ergodic rates, each codeword must span a large number of fading states (i.e., channel matrices). 
Therefore, the decoding order cannot depend on the realization of the channel matrices, but only on their statistics, i.e., ultimately on 
the distance dependent path strengths. Assuming successful decoding and interference cancellation
for the first $\ell - 1$ SIC stages, the signal at the input of the $\ell$-th decoding stage is given by
\begin{align}
{\Ym}^{\rm SIC}_{k,\ell} 
&= \sum_{j= \ell}^{L} \hv_{k,{\pi(j)} }\sqrt{\beta r_{k,{\pi(j)} }^{-\eta}} \, \underline{\xv}_{\pi(j)}  +  \sum_{j= L+1}^{\infty} \hv_{k,j}\sqrt{\beta r_{k,j}^{-\eta}} 
\underline{\xv}_{j}  \quad \ell \leq L
\end{align} 

Denoting by $\Hm_{k,\pi(\ell : L)}$  the channel matrix restricted to ENs with indices $\pi(\ell), \pi(\ell+1), \dots, \pi(L)$ and letting 
$\Hm_{k,\pi(\ell : L)}\ssymbol{2}$ be the corresponding pseudo-inverse, the ZF beamforming vector for the $\ell$-th SIC stage 
is given by the first column of pseudo-inverse, denoted by $[\Hm_{k,\pi(\ell : L)}\ssymbol{2}]_{:,1}$, normalized to have unit norm. Letting 
$\vv_{k,\ell}$ denote such vector, the decoder input after ZF beamforming is given by 
\begin{align}
\tilde{\underline{\yv}}_{k,\ell}^{\rm SIC} 
&=  \vv_{k,\ell}^\herm {\Ym}^{\rm SIC}_{k,\ell} \\
&=  \vv_{k,\ell}^\herm \hv_{k,\pi(\ell)}\sqrt{\beta r_{k,\pi(\ell)}^{-\eta}} \, \underline{\xv}_{\pi(\ell)}+ \sum_{j= L+1}^{\infty} \vv_{k,\ell}^{\herm} \hv_{k,j}\sqrt{\beta r_{k,j}^{-\eta}} \, \underline{\xv}_{j} .
\end{align}
Consequently, the $\ell$-th decoder SIR is given by 
\begin{eqnarray} \label{sir-sic}
\SIR_{k,\ell}^{\rm SIC} &=& \frac{ r_{k,\pi(\ell)}^{-\eta} \left | [\Hm_{k,\pi(\ell : L)}\ssymbol{2}]_{:,1} \right |^{-2}}{\sum_{j = L+1}^{\infty} r_{k,j}^{-\eta}   | \tilde{h}_{k,j} |^2 }.
\end{eqnarray}
Since  $\vv_{k,l}$ is a unit vector that is independent  of $\{ \hv_{k,j} : j  > L \}$, as before 
it follows that $\tilde{h}_{k,j} \sim \mathcal{CN}(0,1), \forall j > L$. Also, similarly as before, we have that 
the useful signal coefficient 
$\left | [\Hm_{k,\pi(\ell : L)}\ssymbol{2}]_{:,1} \right |^{-2} = \mathcal{X}_{2(n_r-L+\ell)}$ is Chi-square distributed with $2(n_r-L+\ell)$ degrees of freedom \and mean 
$n_r - L + \ell$.  
It follows that the ergodic spectral efficiency achievable at the $\ell$-th decoding stage is given by 
\begin{equation} \label{cazzo1sic}
{C}_{k,\ell}(\rho_{\pi(\ell)}) = \EE \left [ \left . \log \left (1+    \frac{ r_{k,\pi(\ell)}^{-\eta} \mathcal{X}_{2(n_r-L+\ell)} }{\sum_{j= L+1}^{\infty} r_{k,j}^{-\eta}  | \tilde{h}_{k,j} |^2 } \right )  \right  |\Phi \right ],  
\end{equation}
that, using Jensen's inequality, can be lower bounded by 
\begin{equation} \label{cazzo1sic-lb}
{C}_{k,\ell}(\rho_{\pi(\ell)}) \geq \EE \left [ \left . \log \left (1+    \rho_{k,\pi(\ell)} \Xc_{2(n_r-L+\ell)} \right )  \right  |\Phi \right ],  
\end{equation}
where $\rho_{k,\ell}$ is defined in (\ref{eq:actual_SIR}).  
Operating as before, we obtain the corresponding quasi-lower bound on the ergodic spectral efficiency  at the $\ell$-th decoding stage as
\begin{equation} \label{Clb-sic}
{C}_{k,\ell}^{\rm qlb}(\tilde{\rho}_{k,\pi(\ell)} ) = \EE \left[ \left . \log \left(  1+  \tilde{\rho}_{k,\pi(\ell)} \mathcal{X}_{2(n_r-L+\ell)} \right ) \right | \Phi \right], 
\end{equation}
where $\tilde{\rho}_{k,\ell}$ is defined in (\ref{ziopippa}). 

Next, we consider the outage probability of the PZF-SIC decoder.  Notice that (\ref{cazzo1sic-lb}) is 
the ergodic achievable rate at the $\ell$-th stage assuming a genie-aided SIC decoder that has already removed
the codewords $\underline{\xv}_{\pi(1)}, \ldots, \underline{\xv}_{\pi(\ell-1)}$. Then,  
we follow an argument similar to the one provided in \cite{rimoldi1996rate}, 
to show that for SIC, the error probability under genie-aided cancellation is the same as the actual error probability.  
It is important to notice that this statement holds only for the overall probability of error, and not for the individual probability of error of each 
decoding stage.  We define the outage set $\Ac$ as the set of all geometry configurations $\Phi$ 
for which the PZF-SIC fails with very high probability (w.h.p). 
For any $\Phi \in \Ac$ there exists $1 \leq \ell^* \leq L$ such that 
\begin{equation} \label{outage-ell}
{C}_{k,\ell}(\rho_{k,\pi(\ell)}) > R, \;\;\; \mbox{for} \; \ell =1,\ldots, \ell^*-1, \;\;\;\; \mbox{and} \;\;\; {C}_{k,\ell^*}(\rho_{k,\pi(\ell^*)} ) \leq  R.
\end{equation}
In other words, $\ell^*$ is the index of the first stage (dependent on $\Phi$) for which a decoding error occurs w.h.p..
Now, consider the set $\Bc$ of geometry configurations $\Phi$ such that  $\min_{\ell \in [L]}  {C}_{k,\ell}(\rho_{k,\pi(\ell)}) \leq R$. 
Since for any $\Phi \in \Ac$ it is always true that  $\min_{\ell \in [L]}  {C}_{k,\ell}(\rho_{k,\pi(\ell)} ) \leq {C}_{k,\ell^*}(\rho_{k,\pi(\ell^*)} ) \leq R$, then 
$\Ac \subseteq \Bc$. On the other hand, let $\hat{\ell}$ denote the index achieving the minimum 
of $\{{C}_{k,\ell}(\rho_{k,\pi(\ell)} ) : \ell \in [L]\}$. For any $\Phi \in \Bc$ we have 
${C}_{k,\hat{\ell}}(\rho_{k,\pi(\hat{\ell})} ) \leq R$. Therefore, there must exist some 
$\ell^* \leq \hat{\ell}$ for which the condition (\ref{outage-ell}) is verified (in fact, it has to be 
some $\ell^* \in [1: \hat{\ell}]$ including $\hat{\ell}$ itself). 
Hence, we have $\Bc \subseteq \Ac$. Thus, we conclude that 
$\Ac = \Bc$, and the outage probability of the PZF-SIC decoder is given by 
$P_{{\rm out},k}(R) =  \PP\left ( \min_{\ell \in [L]} {C}_{k,\ell}(\rho_{k,\pi(\ell)}) \leq R \right )$. Using the same argument made for 
PZF in the previous section, we replace ${C}_{k,\ell}(\rho_{k,\pi(\ell)})$ with its quasi-lower bound approximation (\ref{Clb-sic}) and obtain
a quasi-upper bound to the outage probability of the PZF-SIC as
\begin{align} \label{outage-min}
P_{{\rm out},k}(R) &  \lessapprox  \PP\left ( \min_{\ell} {C}_{k,\ell}^{\rm qlb}(\tilde{\rho}_{k,\pi(\ell)}) \leq R \right ). 
\end{align}
A problem that remains to be addressed is how to choose the optimal decoding order $\pi \in \Pi$ that minimizes the outage probability. 
It is immediate to verify from (\ref{Clb-sic}) that the functions ${C}_{k,\ell}^{\rm qlb}( \rho )$ for $\rho \in \RR_+$ 
are monotonically increasing and that, for any $\rho \in \RR_+$, they satisfy the dominance condition
\begin{equation} \label{dominance}
{C}_{k,1}^{\rm qlb}( \rho ) \leq {C}_{k,2}^{\rm qlb}( \rho ) \leq \cdots \leq {C}_{k,L}^{\rm qlb}( \rho ), 
\end{equation}
due to the order of the chi-squared variable in (\ref{Clb-sic}) that increases with $\ell$.  Then, we have the following result, proved in Appendix \ref{perm-theorem-proof}:

\begin{theorem}  \label{perm-theorem}
Let $f_1(x), \ldots, f_L(x)$ be a collection of monotonically non-decreasing functions on $\RR$ such that, for all $x \in \RR$, 
$f_1(x) \leq f_2(x) \leq \cdots \leq f_L(x)$. Then, for any $L$-tuple of values $x_1 \geq  \cdots \geq x_L$, 
\begin{equation} \label{opt-perm} 
\max_{\pi \in \Pi} \min_{\ell \in [L]} f_\ell(x_{\pi(\ell)} ) = \min_{\ell \in [L]} f_\ell(x_\ell), 
\end{equation}
i.e., the minimum of $f_\ell(x_{\pi(\ell)})$ is maximized by the identity $\pi(\ell) = \ell$ (although this may not be the unique solution). 
\hfill $\square$
\end{theorem}

Notice that for any given realization of the geometry $\Phi$, the sequence $\tilde{\rho}_{k,1}, \tilde{\rho} _{k, 2}, \ldots, \tilde{\rho} _{k, L}$ is monotonically decreasing. 
Therefore, the optimal PZF-SIC decoding order consists of decoding the EN signals in path strength order (strongest first, $L$-th strongest last, 
and treat all the others as noise). 

\begin{figure}
	\centering
	\includegraphics[scale=0.25]{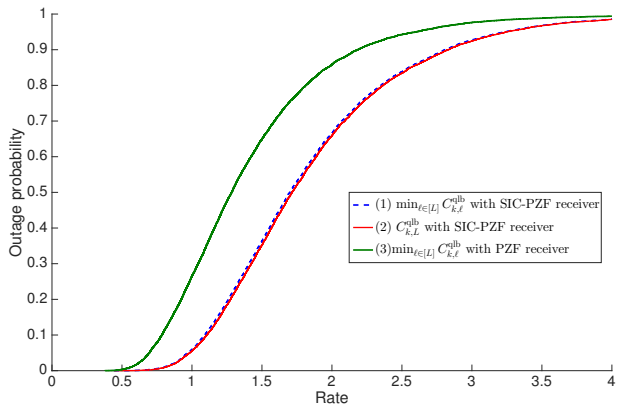}
	\caption{\small Outage probability versus delivery rate with $n_r= 8$. }
	\label{fig:prob_out_comp}
\end{figure} 

In Fig. \ref{fig:prob_out_comp}, the outage probability of the worst stream for cases with PZF and SIC-PZF are plotted. In PZF receiver, 
the last stream ($L$-th one) has the minimum ergodic spectral efficiency and defines the outage probability of PZF (as already written in (\ref{outagep1})). 
However, for the PZF-SIC receiver the last stream is not necessarily the minimum for any realization of $\Phi$. 
The curves $(1)$ and $(2)$ in Fig. \ref{fig:prob_out_comp} show the CDFs 
 $\PP\left ( \min_{\ell \in [L]} {C}_{k,\ell}^{\rm qlb}(\rho_{k,\ell}) \leq R \right )$ and $\PP\left ( {C}_{k,L}^{\rm qlb}(\rho_{k,L}) \leq R \right )$ for SIC-PZF. 
We notice that these two CDFs are almost identical. Therefore, for the sake of analytical tractability, we shall further relax our 
outage probability approximation for the PZF-SIC receiver to 
\begin{align} \label{outage-min-approx}
P_{{\rm out},k}(R) &  \lessapprox  \PP\left (  {C}_{k,L}^{\rm qlb}(\tilde{\rho}_{k,L}) \leq R \right ), 
\end{align}
which can be obtained analytically by adapting the formulas of Section \ref{outage-section}.

We conclude this section by providing a tight approximation of the average (over the geometry) ergodic spectral efficiency of the PZF-SIC decoder. 
From what said before, it follows that 
\begin{eqnarray} 
\overline{C}^{\rm qlb}_{k,{\rm PZF-SIC}} & = & \EE \left [ \min_{\ell \in [L]} {C}_{k,\ell}^{\rm qlb}(\tilde{\rho}_{k,\ell} ) \right ] \nonumber \\
& \approx & \EE \left [ {C}_{k,L}^{\rm qlb}(\tilde{\rho}_{k,L} ) \right ],  \label{average-ergodic-sic}
\end{eqnarray}
where the last line follows again by replacing the minimum with the $L$ term. Using similar steps as done before for the PZF case with $\ell = L$, we obtain
%
\begin{align}
\label{eq: sp_worst_sic}
\overline{C}_{k,{\rm PZF-SIC}}^{\rm qlb}  &= \sum_{ m= 0}^{n_r} \frac{\log_2(e) }{m+L} \Hypergeometric{2}{1}{1,L}{m+L+1}{1-\frac{2}{\eta-2}}.
\end{align}

\section{ Numerical results}
\label{sec:NumericalResults}

In this section we provide some numerical results illustrating the behavior and the performance of the proposed  content delivery wireless caching network.  In order to generate the simulation results, the locations of ENs are distributed according to a homogeneous PPP restricted to a disk radius $R_{\rm area}= 3$ km with density $\lambda =8 \, \mathrm{ENs/km}^2$. The number of ENs is a Possion random variable with mean 
$\pi R_{\rm area}^2 \lambda$ and the EN positions are generated  independently with uniform probability over the disk. 
We considered realistic values of the pathloss exponent $\eta = 3.75$ and the number of antennas $n_r=8,16$ at the user receivers. 

\begin{figure}
	\centering
	\includegraphics[scale=0.25]{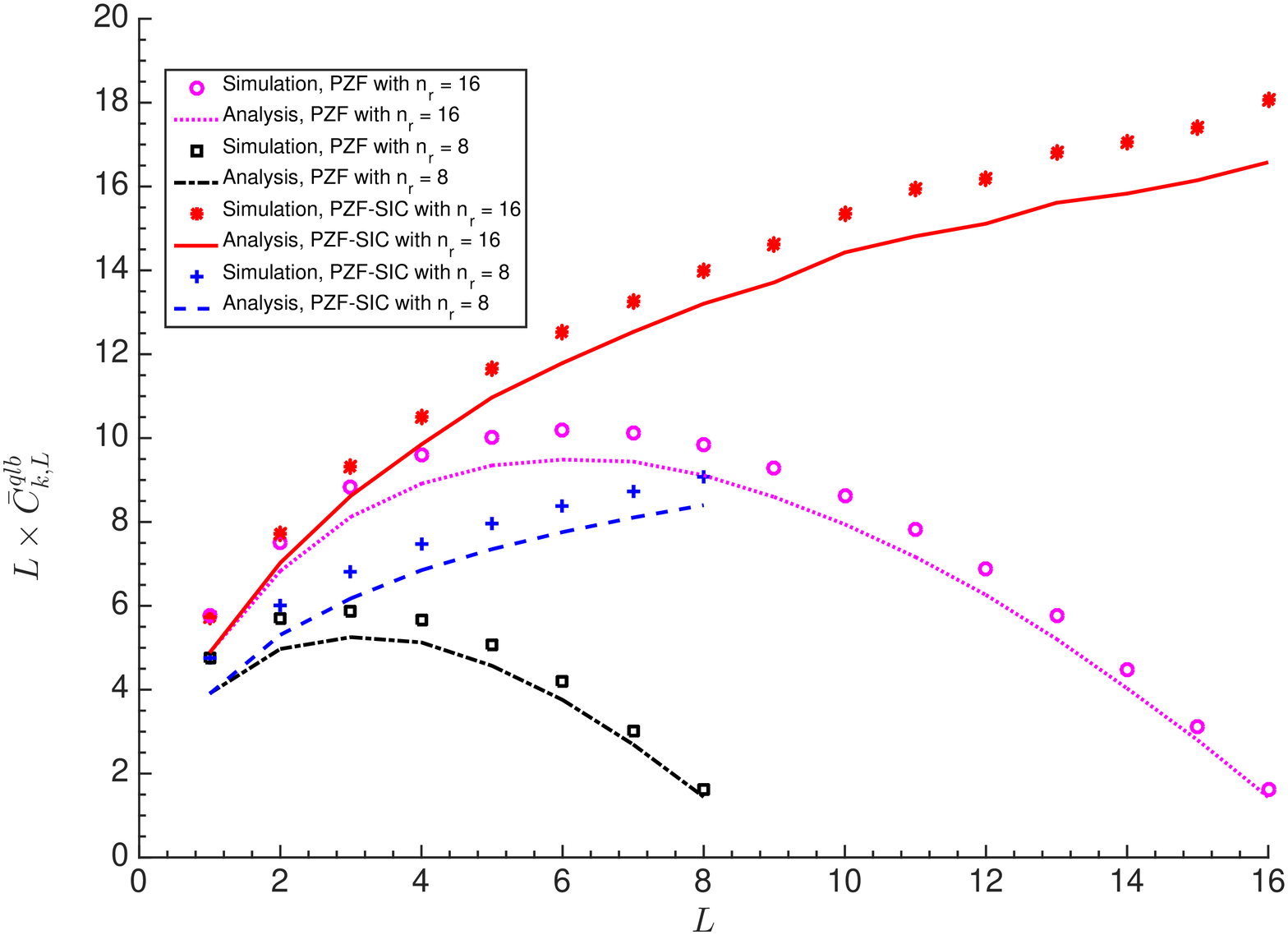}
	\caption{\small  $L \times \overline{C}^{\rm lb}_{k,{\rm PZF}}$ and $L \times \overline{C}^{\rm qlb}_{k,{\rm PZF-SIC}}$versus $L \leq n_r$ for a typical user with $n_r = 8, 16$.}
	\label{fig:capacity}
\end{figure}

Fig. \ref{fig:capacity} compares, as a function of $n_r$,  $L \times \overline{C}^{\rm lb}_{k,{\rm PZF}}$ (cf. (\ref{eq:lb_pzf})) and $L \times \overline{C}^{\rm qlb}_{k,{\rm PZF-SIC}}$(cf. (\ref{eq: sp_worst_sic}))
versus $L$ for PZF and PZF-SIC receiver, which serves for the system designer a  quick way to 
choose the best macro-diversity order $L$ in order to minimize the delivery latency. 
With PZF, the trade-off between PHY rate and macro-diversity order $L$ is evident. In fact, the product 
$L \times \overline{C}^{\rm lb}_{k,{\rm PZF}}$ increases for small $L$, reaches a maximum at $L = 3$ for $n_r = 8$ and
at $L = 6$ for $n_r = 16$,  and then decreases, because the PHY rate at which the typical user can decode the $L$-th strongest EN 
decreases faster than the increase in $L$. In contrast, with PZF-SIC, the product  $L \times \overline{C}^{\rm qlb}_{k,{\rm PZF-SIC}}$ monotonically 
increases with $L \in \{1, \ldots, n_r\}$ showing that, at least for these system parameters, the best macro-diversity order is $L = n_r$.

\begin{figure}
	\centering
	\includegraphics[scale=0.25]{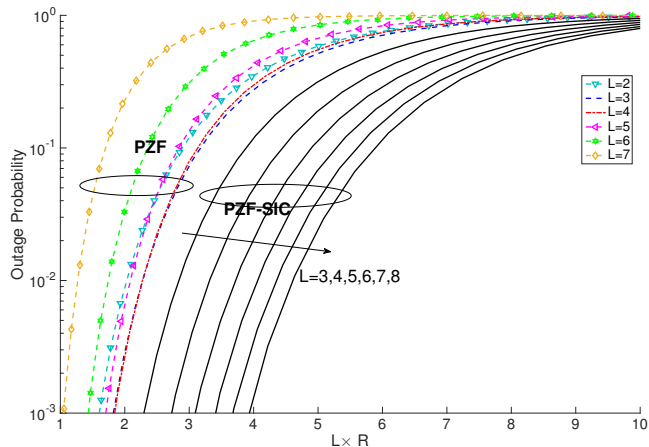}
	\caption{\small Outage probability versus $L\times R$ with $n_r=8$.}
	\label{fig:prob_out_nr8}
\end{figure}

\begin{figure}
	\centering
	\includegraphics[scale=0.25]{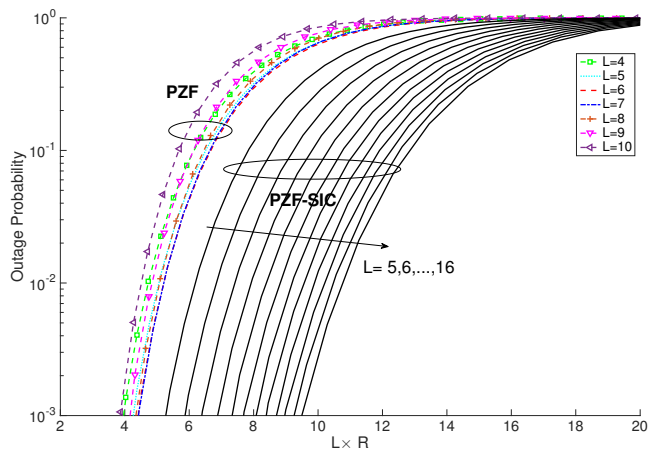}
	\caption{\small Outage probability versus $L\times R$  with $n_r= 16$ .}
	\label{fig:prob_out_nr16}
\end{figure} 

Figs. \ref{fig:prob_out_nr8} and \ref{fig:prob_out_nr16} illustrate the outage probability versus
$L \times R$ with PZF and PZF-SIC receiver, $n_r =8$ and $n_r=16$ , respectively. For better exposition and  simplicity of illustration, we only bring the plots for some macro-diversity orders whose outage probability are the lowest. The missing marco diversity orders in these two figures have higher outage probability than the illustrated plots, so they do not play a role to find the best trade-off between the latency and the macro diversity order.
The delivery latency varies with $L \times R$ in an inverse manner and therefore moving to the right of these curves yields shorter delivery latency at higher outage probability (i.e., 
less and less users will be able to retrieve their requests, but those who can, receive it faster). 
In Fig. \ref{fig:prob_out_nr8},  by using PZF alone the macro-diversity order $L = 3,4$,  achieves better overall product $L \times R$ and therefore  lower delivery latency. 
In PZF-SIC, by increasing $L$ the curves move to right side, indicating that for given target outage probability 
the delivery latency decreases by increasing the macro-diversity order $L$, confirming the behavior already observed for the average ergodic rate
in Fig. \ref{fig:capacity}. Similarly for $n_r=16$, with  PZF receiver macro-diversity $L=6,7$ yield uniformly the best trade-off between 
latency and the outage probability while  for PZF-SIC the best tradeoff is obtained by increasing $L$.

\section{Conclusion}

In this paper, the coded caching paradigm is applied to a multipoint multicast system, somehow reminiscent of the current eMBMS for media broadcasting.
In fact, the proposed system can be seen as a proposal to support on-demand multimedia delivery using the same  principles of eMBMS, i.e., 
simultaneous transmission of a common multicast stream from multiple infrastructure nodes, referred to here as Edge Nodes (ENs). 
The key ingredients of our scheme are coded caching as proposed for the single bottleneck network, able to turns the individual demands
into a single coded multicast stream in an information-theoretic optimal way, and MDS-coded multipoint multicast. 
The MDS code allows each user to retrieve the whole multicast message by decoding just $L$ out of all possible EN transmissions. 
In this way, each user can select its ``best'' $L$ ENs in a completely decentralized user-centric  way. In our case, since the pathloss depends on distance, the 
best $L$ ENs are the $L$ closets ones.  

Overall, the proposed system results in full spatial scalability, since  the per-user throughput (resp., per-request delivery latency) 
does not vanish (resp., does not diverges to infinity) as the system coverage area grows without bound with constant user density, as long as also
the density of the ENs is constant.   We derived analytical expressions for the outage probability with respect 
to the random network geometry and the average ergodic spectral efficiency for a typical user, where the averaging is also 
with respect to the network geometry. Our analysis shows that a rule of thumb to determine the optimal macro-diversity order $L$ 
in the case of linear  Partial Zero-Forcing (PZF) strategy at the users' receiver is to set $L$ equal to half of the number of 
antennas $n_r$ at the user receivers. In contrast, when PZF is augmented by Successive Interference Cancellation (SIC) 
in the order of the strongest EN (first) to the $L$-th strongest (last), letting $L = n_r$ provides the best performance.

\appendices


\section{Proof of (\ref{eq:Laplace_ell}),  (\ref{eq:Laplace_L})}
\label{ap:Laplace_interference}

\textbf{Case 1:} For $\ell<L$ 
\begin{equation}
\mathcal{L}_{1 /\tilde{\rho}_{k,\ell}}(s) =  \int_{0}^{\infty} dv \int_{0}^{v}  e^{-s \frac{u^{\eta} }{v^{\eta-2}} \frac{2 \pi \lambda}{\eta-2}} f_{r_{k,\ell}, r_{k,L}}(u, v) d u.
\end{equation}
The joint PDF of $f_{r_{k,\ell}, r_{k,L}}(u, v)$ is given by (\ref{eq:jointrr}) with $n\leftarrow L$, $\ell \leftarrow \ell$. By replacing the expression for the joint PDF, we have
\begin{align}
\mathcal{L}_{1 /\tilde{\rho}_{k,\ell}}(s) =  \int_{0}^{\infty} dv \int_{0}^{v} & e^{-s \frac{u^{\eta} }{v^{\eta-2}} \frac{2 \pi \lambda}{\eta-2}} \frac{4 (\pi \lambda)^{L}}{(L-\ell-1)! \, (\ell - 1)!} \nonumber  \\
&  \cdot (v^2-u^2)^{L-\ell-1} v u^{2\ell-1} e^{-\pi \lambda v^2}d u,
\end{align}
where after invoking the binomial expansion for $(v^2-u^2)^{L-\ell-1}$, we obtain
\begin{align}
\mathcal{L}_{1 /\tilde{\rho}_{k,\ell}}(s) &= \sum_{n=0}^{L-\ell-1} \frac{4 (\pi \lambda)^{L}}{(L-\ell-1)! \, (\ell - 1)!} 
{{L-\ell-1} \choose {n}} (-1)^n \nonumber \\
&\quad .\int_{0}^{\infty} v^{2(L-\ell-1-n)+1}e^{-\pi \lambda v^2}dv  \int_{0}^{v}e^{-s \frac{2 \pi \lambda}{\eta-2}\frac{u^{\eta}}{v^{\eta-2}}} u^{2\ell-1+2n}d u.
\end{align}
First, we apply a change of variable $u^{\eta} \rightarrow x$ and then utilize \cite[3.351.1]{jeffrey2007table} to solve the inner integral. After simplification, we have 
\begin{align}
\mathcal{L}_{1 /\tilde{\rho}_{k,\ell}}(s) &=  \sum_{n=0}^{L-\ell-1} \frac{4 (\pi \lambda)^{L}}{(L-\ell-1)! \, (\ell - 1)!} 
{{L-\ell-1} \choose {n}} (-1)^n \int_{0}^{\infty} v^{\nu+1 }e^{-\pi \lambda v^2}  \bar{\Gamma}(2\frac{n+\ell}{\eta},s \alpha_1 v^2) dv,
\end{align}
where $\nu = 2((L-\ell-1-n)+(\eta-2)\frac{n+\ell}{\eta})$ and $\alpha_1 = \frac{2\pi \lambda}{\eta-2}$.
Then, we invoke the change the variable $v^2 \rightarrow x$ once more and then utilize the identity \cite[6.455.2]{jeffrey2007table} to obtain following result
\begin{align}
\mathcal{L}_{1 /\tilde{\rho}_{k,\ell}}(s) = & \sum_{n=0}^{L-\ell-1} \frac{(-1)^n \Gamma(L)}{(L-\ell-1)! \, (\ell - 1)! \, (n+\ell)} 
{{L-\ell-1} \choose {n}} \nonumber \\
&\cdot (1+\alpha_2 s)^{-L}\Hypergeometric{2}{1}{1,L}{\eta'+1}{\frac{\alpha_2 s}{1+\alpha_2 s}}
\end{align}
where $\alpha_2 = \frac{2}{\eta-2}$, $\eta' = 2\frac{n+\ell}{\eta}$. Finally, by using the formula \cite[9.131.1]{jeffrey2007table} the Laplace transform can be rewritten as in (\ref{eq:Laplace_ell}).

\textbf{Case 2:} For $\ell=L$
\begin{align}
\mathcal{L}_{1 /\tilde{\rho}_{k,L}}(s) = \int_{0}^{\infty} e^{-s  v^2 \frac{2 \pi \lambda}{\eta-2}} f_{ r_{k,L}}( v) d v,
\end{align}
The marginal PDF $f_{r_{k,L}}(v)$ is given by (\ref{eq:marginal_pdf}) with $n\leftarrow L$. After replacing the expression for the marginal PDF, we have 
\begin{align}
\mathcal{L}_{1 /\tilde{\rho}_{k,L}}(s) = \int_{0}^{\infty} e^{-s  v^2 \frac{2 \pi \lambda}{\eta-2}} \frac{2(\pi \lambda)^{L}}{(L-1)! } v^{2L-1} e^{-\pi \lambda v^2} d v,
\end{align}
To solve this integral first we apply the variable change $v^2 \rightarrow x$ and then use \cite[3.351.3]{jeffrey2007table}, which gives (\ref{eq:Laplace_L}).

\section{Proof of (\ref{eq:CDF_SIR_ell}) and (\ref{eq:CDF_SIR_L})}
\label{ap:proof_prop2}

\textbf{Case 1:} For $\ell<L$
\begin{align}
F_{\widetilde{\SIR}_ {{k,\ell}} }(\gamma) = 1- \int_{0}^{\infty} \int_{0}^{v} (1-F_{\widetilde{\SIR}_{{k,\ell}} } (\gamma| r_{k,\ell}, r_{k,L}) )f_{r_{k,\ell}, r_{k,L}}(u, v) dudv. \label{eq:int_3}
\end{align}
By direct substitution of the expression (\ref{eq:chi_sir})  into (\ref{eq:int_3}), we have
\begin{align}
F_{\widetilde{\SIR}_ {{k,\ell}} }(\gamma) =1-& \int_{0}^{\infty} \int_{0}^{v} 
\frac{\Gamma \left(n_r-L+1,  \gamma\frac{2 \pi \lambda}{\eta-2}\frac{u^{\eta}}{v^{\eta-2}}\right )}{\Gamma{(n_r-L+1)}} \nonumber \\
&.\frac{4 (\pi \lambda)^{L}}{(L-\ell-1)! \, (\ell - 1)!}(v^2-u^2)^{L-\ell-1} v u^{2\ell-1} e^{-\pi \lambda v^2} du dv.
\end{align}
Note that by using \cite[8.352.2]{jeffrey2007table} we can expand the incomplete gamma function and then we have
\begin{align}
F_{\widetilde{\SIR}_ {{k,\ell}} }(\gamma) = 1-&\sum_{m=0}^{n_r-L}\frac{4 (\pi \lambda)^{L}}{(L-\ell-1)! \, (\ell - 1)!\, m!} \nonumber \\
&  \cdot \int_{0}^{\infty} \int_{0}^{v}  ( \gamma\frac{2 \pi \lambda}{\eta-2}\frac{u^{\eta}}{v^{\eta-2}})^m e^{-  \gamma\frac{2 \pi \lambda}{\eta-2}\frac{u^{\eta}}{v^{\eta-2}}} (v^2-u^2)^{L-\ell-1} v u^{2\ell-1} e^{-\pi \lambda v^2} du dv. 
\end{align}
After expanding binomial, we have 
\begin{align}
F_{\widetilde{\SIR}_ {{k,\ell}} }(\gamma)  
&= 1-\sum_{n=0}^{L-\ell-1}\sum_{m=0}^{n_r-L}\frac{4 (\pi \lambda)^{L}(-1)^n}{(L-\ell-1)! \, (\ell - 1)!\, m!}(\gamma \frac{2 \pi \lambda}{\eta-2})^m {{L-\ell-1} \choose {n}} \nonumber\\
&\quad \cdot \int_{0}^{\infty}  v^{-m(\eta-2)+2(L-\ell-1-n)+1} e^{-\pi \lambda v^2}  \int_{0}^{v}
e^{-  \gamma\frac{2 \pi \lambda}{\eta-2}\frac{u^{\eta}}{v^{\eta-2}}} 
  u^{2n+2\ell-1+m\eta} du dv
\end{align}
To solve the inner integral first we change the variable $u^{\eta} \rightarrow x$ and then use \cite[3.351]{jeffrey2007table},
\begin{align}
F_{\widetilde{\SIR}_ {{k,\ell}} }(\gamma) &=  1-\sum_{n=0}^{L-\ell-1}\sum_{m=0}^{n_r-L}\frac{4 (\pi \lambda)^{L}(-1)^n}{(L-\ell-1)! \, (\ell - 1)!\, m!} {{L-\ell-1} \choose {n}}(\gamma \frac{2 \pi \lambda}{\eta-2})^{-\eta'} \nonumber \\
&\quad \cdot \int_{0}^{\infty} v^{\nu+1} \bar{\Gamma}(\eta'+m, \gamma\alpha_1 v^2)e^{-\pi \lambda v^2} dv,
\end{align}
where $\nu =(\eta-2)\eta'+2(L-\ell-1-n)$ and $\alpha_2 =  \frac{2}{\eta-2}$, $\eta' = 2\frac{n+\ell}{\eta}$

In the next step to solve the remaining outer integral, first we change the variable $v^2 \rightarrow x$ and then use \cite[6.455.2]{jeffrey2007table} to obtain following result
\begin{align}
F_{\widetilde{\SIR}_ {{k,\ell}} }(\gamma) &=1-\sum_{n=0}^{L-\ell-1}\sum_{m=0}^{n_r-L} \frac{2(-1)^n\Gamma(m+L)}{(L-\ell-1)! \, (\ell - 1)!\, m! \, \eta \, (\eta'+m)}{{L-\ell-1} \choose {n}} \nonumber\\
&\quad \cdot \frac{( \alpha_2\gamma)^m}{(\alpha_2\gamma+1)^{m+L}} 
\Hypergeometric{2}{1}{1,L+m}{\eta'+m+1}{\frac{\alpha_2 \gamma}{1+\alpha_2 \gamma}}. \nonumber
\end{align}
where $\alpha_2 = \frac{2}{\eta-2}$. Finally, we use the formula \cite[9.131.1]{jeffrey2007table} to rewrite the preceding expression as (\ref{eq:CDF_SIR_ell}).

\textbf{Case 2:} For $\ell=L$

As previous case, the complement CDF of $\SIR$ can be written as follows
\begin{equation}
\begin{split}
F_{\widetilde{\SIR}_{k,L}} (\gamma)=&1- \int_{0 }^{\infty}  \frac{\Gamma \left(n_r-L+1, v^2 \gamma \frac{2 \pi \lambda }{\eta-2}\right) }{\Gamma(n_r-L+1)} \frac{2 (\pi \lambda)^{L}}{(L-1)! }v^{2L-1} e^{-\pi \lambda v^2} dv.\\
\end{split}
\label{eq:int_fisrt}
\end{equation}
Note that by using \cite[8.352.2]{jeffrey2007table} we can expand the incomplete gamma function and then then the preceding integral can be written as follows
\begin{equation}
\begin{split}
&F_{\widetilde{\SIR}_{k,L}} (\gamma) = 1-\sum_{m=0}^{n_r-L} \frac{2 (\pi \lambda)^{L}}{(L-1)!\, m!} (\gamma \alpha_1)^m \int_{ 0}^{\infty}v^{2m+2L-1} e^{-(\pi \lambda+\alpha_1 \gamma) v^2}dv,
\end{split}
\end{equation}
where $\alpha_1 = \frac{2\pi \lambda}{\eta-2}$. After solving the integral by using \cite[3.351.3]{jeffrey2007table} and simplify the result, the final expression can be written as (\ref{eq:CDF_SIR_L}).

\section{Proof of (\ref{eq:Avg Seff ell expr}) and (\ref{eq: sp_worst})} 
\label{ap:proof_prop3}
The average spectral efficiency can be computed by a integral which we provided in  (\ref{eq: avg_spec}).

\textbf{Case 1:} For $\ell<L$
\begin{equation}
\overline{C}_{k,\ell}^{\rm lb}=\int_{0}^{\infty} 1-F_{\widetilde{\SIR}_{k,\ell}}(2^{\gamma}-1)d\gamma 
\end{equation}
By substituting the complement CDF of $\SIR$ from (\ref{eq:CDF_SIR_ell}) into the preceding expression, we have
\begin{align}
 \overline{C}_{k,\ell}^{\rm lb} = &\sum_{ m= 0}^{n_r-L}\sum_{n= 0}^{L-\ell-1}  \frac{2 (-1)^n }{m\, ! \, (L-\ell-1)\, ! \, (\ell-1)\, !  }   \frac{\Gamma(m+L)}{\eta \, (\eta'+m)}  {{L-\ell-1} \choose {n}} \\ 
 & \cdot  \int_{0}^{\infty}(\alpha_2 (2^{\gamma}-1))^m  
\Hypergeometric{2}{1}{L+m,m+\eta'}{m+\eta'+1}{-(2^{\gamma}-1) \alpha_2} d\gamma. \nonumber 
\end{align}

By changing $2^{\gamma}-1  \rightarrow x$ and  using of \cite[9.34.7]{jeffrey2007table} , the hypergeometric function can be expressed in terms of the Meijer-G function as follows
\begin{align}
 \overline{C}_{k,\ell}^{\rm lb} = &\sum_{ m= 0}^{n_r-L}\sum_{n= 0}^{L-\ell-1}  \frac{2 (-1)^n }{m\, ! \, (L-\ell-1)\, ! \, (\ell-1)\, ! \, \eta }  {{L-\ell-1} \choose {n}} \log_2(e)\\
 &\cdot \int_{0}^{\infty}(\alpha_2 x)^{m+1} (x+1)^{-1}  
   G_{2,2}^{1,2} \left(\begin{matrix} -(m+L),-m-\eta' \\
  -1,-(m+1+\eta')
 \end{matrix} \bigg| \alpha_2 x\right)  dx\nonumber 
\end{align}
and the preceding  integral has the explicit solution in \cite[7.811.5]{jeffrey2007table} and is provided the final expression in (\ref{eq:Avg Seff ell expr}).

\textbf{Case 2:} For $\ell=L$
\begin{eqnarray}
\overline{C}_{k,L}^{\rm lb}=\int_{0}^{\infty} F^c_{\widetilde{\SIR}_{k,\ell}}(2^{\gamma}-1)d\gamma 
\end{eqnarray}
by changing varibale to $2^{\gamma}\rightarrow e^{\gamma}$, then the integral is rewritten as
\begin{equation}
\overline{C}_{k,L}^{\rm lb}= \sum_{m=0}^{n_r-L} \frac{(m+L-1)!}{m\, !\, (L-1)\,!\,}   \frac{\log_2 e}{y^L}  \int_{0}^{\infty} \frac{e^{-\gamma L}(1-e^{-\gamma})^m}{(1+ \frac{1-y}{y}e^{-\gamma})^{m+L}} d\gamma
\end{equation}
The integral can be solved by using  of \cite[3.312.3]{jeffrey2007table} . The final expression for the integral is
\begin{equation}
\begin{split}
\overline{C}_{k,L}^{\rm lb} =& 
\sum_{m=0}^{n_r-L} \frac{(m+L-1)!}{m\, !\, (L-1)\,!\,} \log_2(e) \frac{\Gamma(L)\Gamma(m+1)}{\Gamma(m+L+1)}   \Hypergeometric{2}{1}{1,L}{m+L+1}{1-\frac{2}{\eta-2}}.  
\end{split}
\label{eq:int_ase}
\end{equation}
After simplification, the final expression for average spectral efficiency for the last sub-stream  will be derived and be given by (\ref{eq: sp_worst}).

\section{Proof of Theorem \ref{perm-theorem}}   \label{perm-theorem-proof}

Consider the $L \times L$ matrix $\Tm$ with $(i,j)$ elements
$T_{i,j} = f_i(x_j), \;\;\; i \in [L], \;\; j \in [L]$.
The elements $T_{i,j}$ form a partially ordered set. By construction, two elements $T_{i,j}$ and $T_{i,j'}$ on the same row $i$ are ordered such that 
$T_{i,j} \geq T_{i,j'}$ for $j \leq j'$, and two elements $T_{i,j}$ and $T_{i',j}$ on the same column $j$ are ordered such that 
$T_{i,j} \geq T_{i',j}$ for $i \geq i'$. 

Consider the Manhattan topology  induced by the squared integer grid of the matrix indices  (similar to the moves of the Rook on a chess board).  
Any two elements in the array are connected by a path formed by a horizontal move followed by a vertical move.
A horizontal move to the left decreases the column index. A vertical move down increases the row index. Hence, 
the above inequalities imply that if two elements $T_{i,j}$ and $T_{i',j'}$ are connected by a path formed by a left horizontal move followed by a down vertical 
move (we write $(i,j) \vdash (\leftarrow, \downarrow) \dashv (i',j')$, we can conclude that $T_{i,j} \leq T_{i',j'}$. In contrast, any move containing 
a right horizontal move or an up vertical move does not lead to a necessary ordering of the elements, since these moves go against the monotonicity 
of the elements in the same row and in the same column given above.  

Consider any permutation $\pi \in \Pi$, represented as $L \times L$ binary matrix $\Pm_\pi$ with a single 1 in each row and column, where 
$\Pim_{\sf id}$ is the identity matrix corresponding to the identity permutation ${\sf id}$. We denote by $\Sc_\pi$ the support of
$\Pm_\pi$, i.e., the list of the positions of its ``ones'', i.e., $\Sc_\pi = \{ (i,j) : [\Pm_\pi]_{i,j} = 1\}$.
The set of values $\{f_\ell (x_{\pi(\ell)}) : \ell \in [L]\}$ are obtained by retaining the elements of $\Tm$ corresponding to the ``ones'' of $\Pm_\pi$. 
In particular, $T_{\min, \pi} = \min \{ T_{i,j} : (i,j) \in \Sc_\pi\}$ is the minimum corresponding to 
permutation $\pi$ appearing in the right-hand side of (\ref{opt-perm}). 
A transposition consists of exchanging two rows of $\Pim_\pi$. 
Assume $(i,j) \in \Sc_\pi$ and $(i',j') \in \Sc_\pi$, and consider the permutation 
$\pi'$ obtained from $\pi$ by transposition of rows $i$ and $i'$. It follows that 
$\Sc_{\pi'}$ contains the same elements of $\Sc_\pi$ with the exception of $(i,j), (i',j')$ which are replaced by $(i,j'), (i',j)$.  
The following statement is immediate  (details are omitted for the sake of space limitation): consider $\Pm_\pi$ and two positions 
$(i,j)$  and $(i',j')$ in its support $\Sc_\pi$ and consider $\Pm_\pi'$ obtained from $\Pm_\pi$ by transposition of rows $i$ and $i'$. 
If $(i,j) \vdash (\leftarrow, \downarrow) \dashv (i',j')$, then $T_{\min, \pi'}  \geq T_{\min, \pi}$. We call such transposition a {\em minimum non-decreasing transposition}.
Notice that any permutation $\pi \neq {\sf id}$ has at least a pair of positions $(i,j)$ and $(i',j')$ in $\Sc_\pi$ 
such that  $(i,j) \vdash (\leftarrow, \downarrow) \dashv (i',j')$. Hence, if $\pi$ achieves the max in (\ref{opt-perm}), then $\pi'$
obtained by minimum non-decreasing transposition that exchanges rows $i$ and $j$ also achieves the same max-min.  
It follows that the Theorem \ref{perm-theorem} is proved by showing that any $\pi$ can be transformed 
into the identity permutation by a sequence of minimum non-decreasing transpositions. 

\begin{figure}
	\centering
	\includegraphics[scale=0.25]{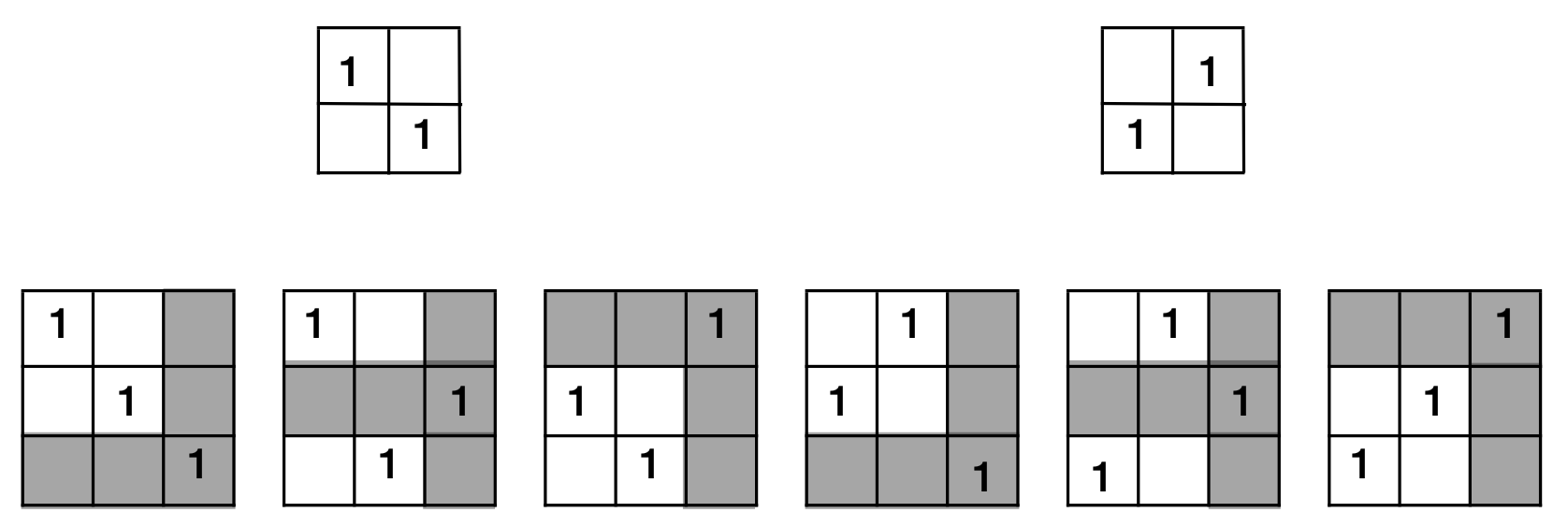}
	\caption{\small Obtaining permutations of order $L = 3$ from the permutations of order $L = 2$ by adding a row and a column (shaded).}
	\label{23example}
\end{figure}

The proof follows by induction. For $L = 2$, 
${\footnotesize \left [ \begin{array}{cc} 0 & 1 \\ 1 & 0 \end{array} \right]}$ can be turned into
${\footnotesize \left [ \begin{array}{cc} 1 & 0 \\ 0 & 1 \end{array} \right]}$ by transposing the rows, and 
obviously $(1,2) \vdash (\leftarrow, \downarrow) \dashv (2,1)$. Therefore, the statement holds for $L = 2$. Suppose that the statement is proved for $L - 1$.
Notice that we can obtain the $L!$  matrices of the order-$L$ permutations from the $(L-1)!$ matrices of the order-$(L-1)$ permutations 
by inserting an additional row in all possible positions and column in last position with a single ``one'' at their intersection, to all
the  order-$(L-1)$ matrices. Fig.~\ref{23example} shows this augmentation process to go from $L = 2$ to $L = 3$. 
For all order-$L$ matrices with the additional ``one'' in position $(L,L)$ there is nothing to prove, since the 
$(L-1) \times (L-1)$ upper left submatrix is an order-$(L-1)$ permutation matrix, that we can transform into the identity by a sequence
of minimum non-decreasing transpositions by the induction assumption. For all order-$L$ matrices with the additional ``one'' in position
$(i,L)$ for some $1 \leq i \leq L-1$, we notice that the last row must contain a ``one'' in position $(L,j)$ for $j < L$. 
Hence, $(i,L) \vdash (\leftarrow, \downarrow) \dashv (L,j)$, i.e., the transposition of rows $L$ and $i$ is minimum non-decreasing. 
After such transposition, we have a one in position $(L,L)$ and the upper left $(L-1) \times (L-1)$ upper left submatrix is an order-$(L-1)$ permutation matrix, 
that can be  transformed into the identity by a sequence
of minimum non-decreasing transpositions. This concludes the proof.  \hfill \QED

{\small
\bibliographystyle{IEEEtran}
\bibliography{references}

\begin{thebibliography}{10}
\providecommand{\url}[1]{#1}
\csname url@samestyle\endcsname
\providecommand{\newblock}{\relax}
\providecommand{\bibinfo}[2]{#2}
\providecommand{\BIBentrySTDinterwordspacing}{\spaceskip=0pt\relax}
\providecommand{\BIBentryALTinterwordstretchfactor}{4}
\providecommand{\BIBentryALTinterwordspacing}{\spaceskip=\fontdimen2\font plus
\BIBentryALTinterwordstretchfactor\fontdimen3\font minus
  \fontdimen4\font\relax}
\providecommand{\BIBforeignlanguage}[2]{{%
\expandafter\ifx\csname l@#1\endcsname\relax
\typeout{** WARNING: IEEEtran.bst: No hyphenation pattern has been}%
\typeout{** loaded for the language `#1'. Using the pattern for}%
\typeout{** the default language instead.}%
\else
\language=\csname l@#1\endcsname
\fi
#2}}
\providecommand{\BIBdecl}{\relax}
\BIBdecl

\bibitem{GSDMC:12}
N.~Golrezaei, K.~Shanmugam, A.~Dimakis, A.~Molisch, and G.~Caire,
  ``Femtocaching: Wireless video content delivery through distributed caching
  helpers,'' in \emph{INFOCOM, 2012 Proceedings IEEE}, March 2012, pp.
  1107--1115.

\bibitem{golrezaei2013femtocaching}
N.~Golrezaei, A.~F. Molisch, A.~G. Dimakis, and G.~Caire, ``Femtocaching and
  device-to-device collaboration: A new architecture for wireless video
  distribution,'' \emph{{IEEE Communications Magazine}}, vol.~51, no.~4, pp.
  142--149, 2013.

\bibitem{liu2016caching}
D.~Liu, B.~Chen, C.~Yang, and A.~F. Molisch, ``Caching at the wireless edge:
  Design aspects, challenges, and future directions,'' \emph{{IEEE
  Communications Magazine}}, vol.~54, no.~9, pp. 22--28, 2016.

\bibitem{paschos2016wireless}
G.~Paschos, E.~Bastug, I.~Land, G.~Caire, and M.~Debbah, ``Wireless caching:
  Technical misconceptions and business barriers,'' \emph{IEEE Communications
  Magazine}, vol.~54, no.~8, pp. 16--22, 2016.

\bibitem{sengupta2016cloud}
A.~Sengupta, R.~Tandon, and O.~Simeone, ``Cloud and cache-aided wireless
  networks: Fundamental latency trade-offs,'' \emph{{Wireless Communications
  (SPAWC)}}, p.~1, 2016.

\bibitem{poularakis2016complexity}
K.~Poularakis and L.~Tassiulas, ``On the complexity of optimal content
  placement in hierarchical caching networks,'' \emph{{IEEE Transactions on
  Communications}}, vol.~64, no.~5, pp. 2092--2103, 2016.

\bibitem{maddah2014fundamental}
M.~A. Maddah-Ali and U.~Niesen, ``Fundamental limits of caching,'' \emph{IEEE
  Transactions on Information Theory}, vol.~60, no.~5, pp. 2856--2867, 2014.

\bibitem{maddah2015decentralized}
------, ``Decentralized coded caching attains order-optimal memory-rate
  tradeoff,'' \emph{IEEE/ACM Transactions On Networking}, vol.~23, no.~4, pp.
  1029--1040, 2015.

\bibitem{yu2017exact}
Q.~Yu, M.~A. Maddah-Ali, and A.~S. Avestimehr, ``The exact rate-memory tradeoff
  for caching with uncoded prefetching,'' in \emph{Information Theory (ISIT),
  2017 IEEE International Symposium on}.\hskip 1em plus 0.5em minus 0.4em\relax
  IEEE, 2017, pp. 1613--1617.

\bibitem{caching-delivery-separation-ISIT}
N.~Naderializadeh, M.~A. Maddah-Ali, and A.~S. Avestimehr, ``On the optimality
  of separation between caching and delivery in general cache networks,'' in
  \emph{Proc. IEEE Int. Symp. Inform. Theory}, June 2017, pp. 1232--1236.

\bibitem{ji2015caching}
M.~Ji, A.~M. Tulino, J.~Llorca, and G.~Caire, ``Caching in combination
  networks,'' in \emph{Signals, Systems and Computers, 2015 49th Asilomar
  Conference on}.\hskip 1em plus 0.5em minus 0.4em\relax IEEE, 2015, pp.
  1269--1273.

\bibitem{zewail2017coded}
A.~A. Zewail and A.~Yener, ``Coded caching for combination networks with
  cache-aided relays,'' in \emph{Information Theory (ISIT), 2017 IEEE
  International Symposium on}.\hskip 1em plus 0.5em minus 0.4em\relax IEEE,
  2017, pp. 2433--2437.

\bibitem{wan2017caching}
K.~Wan, M.~Ji, P.~Piantanida, and D.~Tuninetti, ``Caching in combination
  networks: Novel multicast message generation and delivery by leveraging the
  network topology,'' \emph{arXiv preprint arXiv:1710.06752}, 2017.

\bibitem{wan2017novel}
------, ``Novel outer bounds and inner bounds with uncoded cache placement for
  combination networks with end-user-caches,'' \emph{arXiv preprint
  arXiv:1701.06884}, 2017.

\bibitem{mital2017coded}
N.~Mital, D.~Gunduz, and C.~Ling, ``Coded caching in a multi-server system with
  random topology,'' \emph{arXiv preprint arXiv:1712.00649}, 2017.

\bibitem{ngo2017scalable}
K.-H. Ngo, S.~Yang, and M.~Kobayashi, ``Scalable content delivery with coded
  caching in multi-antenna fading channels,'' \emph{arXiv preprint
  arXiv:1703.06538}, 2017.

\bibitem{simeone2016cloud}
O.~Simeone, A.~Maeder, M.~Peng, O.~Sahin, and W.~Yu, ``Cloud radio access
  network: Virtualizing wireless access for dense heterogeneous systems,''
  \emph{Journal of Communications and Networks}, vol.~18, no.~2, pp. 135--149,
  2016.

\bibitem{C-RAN-compression}
A.~Sanderovich, O.~Somekh, H.~V. Poor, and S.~Shamai, ``Uplink macro diversity
  of limited backhaul cellular network,'' \emph{IEEE Transactions on
  Information Theory}, vol.~55, no.~8, pp. 3457--3478, Aug 2009.

\bibitem{park2014fronthaul}
S.-H. Park, O.~Simeone, O.~Sahin, and S.~S. Shitz, ``Fronthaul compression for
  cloud radio access networks: Signal processing advances inspired by network
  information theory,'' \emph{IEEE Signal Processing Magazine}, vol.~31, no.~6,
  pp. 69--79, 2014.

\bibitem{peng2016fog}
M.~Peng, S.~Yan, K.~Zhang, and C.~Wang, ``Fog-computing-based radio access
  networks: issues and challenges,'' \emph{IEEE Network}, vol.~30, no.~4, pp.
  46--53, 2016.

\bibitem{park2013joint}
S.-H. Park, O.~Simeone, O.~Sahin, and S.~Shamai, ``Joint precoding and
  multivariate backhaul compression for the downlink of cloud radio access
  networks,'' \emph{IEEE Transactions on Signal Processing}, vol.~61, no.~22,
  pp. 5646--5658, 2013.

\bibitem{tandon2016cloud}
R.~Tandon and O.~Simeone, ``Cloud-aided wireless networks with edge caching:
  Fundamental latency trade-offs in fog radio access networks,'' in
  \emph{Information Theory (ISIT), 2016 IEEE International Symposium on}.\hskip
  1em plus 0.5em minus 0.4em\relax IEEE, 2016, pp. 2029--2033.

\bibitem{massiveMIMO-book-Marzetta}
T.~L. Marzetta, E.~G. Larsson, H.~Yang, and H.~Q. Ngo, \emph{Fundamentals of
  massive {MIMO}}.\hskip 1em plus 0.5em minus 0.4em\relax Cambridge, U. K.:
  Cambridge Univ. Press, 2016.

\bibitem{haenggi2012stochastic}
M.~Haenggi, \emph{Stochastic geometry for wireless networks}.\hskip 1em plus
  0.5em minus 0.4em\relax Cambridge University Press, 2012.

\bibitem{distance-distrib-haenggi}
------, ``On distances in uniformly random networks,'' \emph{IEEE Transactions
  on Information Theory}, vol.~51, no.~10, pp. 3584--3586, 20105.

\bibitem{martin2014multi}
F.~J. Martin-Vega, F.~J. Lopez-Martinez, G.~Gomez, and M.~C. Aguayo-Torres,
  ``Multi-user coverage probability of uplink cellular systems: A stochastic
  geometry approach,'' in \emph{Global Communications Conference (GLOBECOM),
  2014 IEEE}.\hskip 1em plus 0.5em minus 0.4em\relax IEEE, 2014, pp.
  3989--3994.

\bibitem{blaszczyszyn2016spatial}
B.~B{\l}aszczyszyn and M.~K. Karray, ``Spatial distribution of the sinr in
  poisson cellular networks with sector antennas,'' \emph{IEEE Transactions on
  Wireless Communications}, vol.~15, no.~1, pp. 581--593, 2016.

\bibitem{abate1995numerical}
J.~Abate and W.~Whitt, ``Numerical inversion of laplace transforms of
  probability distributions,'' \emph{ORSA Journal on computing}, vol.~7, no.~1,
  pp. 36--43, 1995.

\bibitem{o1997euler}
C.~A. O'cinneide, ``Euler summation for fourier series and laplace transform
  inversion,'' \emph{Stochastic Models}, vol.~13, no.~2, pp. 315--337, 1997.

\bibitem{alouini1999capacity}
M.-S. Alouini and A.~J. Goldsmith, ``Capacity of rayleigh fading channels under
  different adaptive transmission and diversity-combining techniques,''
  \emph{{IEEE Transactions on Vehicular Technology}}, vol.~48, no.~4, pp.
  1165--1181, 1999.

\bibitem{MunMorLoz-TWC15}
R.~K. Mungara, D.~Morales-Jim\'{e}nez, and A.~Lozano, ``System-level
  performance of interference alignment,'' \emph{IEEE Transactions on Wireless
  Communications}, vol.~14, no.~2, pp. 1060--1070, Feb. 2015.

\bibitem{rimoldi1996rate}
B.~Rimoldi and R.~Urbanke, ``{A rate-splitting approach to the Gaussian
  multiple-access channel},'' \emph{IEEE Transactions on Information Theory},
  vol.~42, no.~2, pp. 364--375, 1996.

\bibitem{jeffrey2007table}
A.~Jeffrey and D.~Zwillinger, \emph{Table of integrals, series, and
  products}.\hskip 1em plus 0.5em minus 0.4em\relax Academic press, 2007.

\end{thebibliography}
}

%
%

\end{document}